\documentclass[aps,prb,twocolumn,superscriptaddress,floatfix]{revtex4-2}
\usepackage{amsmath,amssymb,bm}
\usepackage{graphicx}
\usepackage{braket}
\usepackage{physics}
\usepackage{color}
\usepackage{tikz}
\usepackage{mathrsfs,euscript}
\usepackage{xfrac}
\usepackage{calc}
\usepackage{graphicx}
\usepackage{xcolor}
\usepackage{xspace}

\definecolor{myblue}{RGB}{210, 235, 255}
\definecolor{myred}{RGB}{255, 220, 230}
\definecolor{mypurple}{RGB}{235, 225, 255}
\definecolor{mygreen}{RGB}{230, 255, 240}

\usetikzlibrary{shapes, shapes.geometric,positioning, calc, decorations.pathreplacing, decorations.markings, decorations.pathmorphing, backgrounds, patterns,  fadings, external}
\usepackage{pgffor}

\tikzset{
    line/.style={line width=0.02cm, black},
    thickline/.style={line width=0.04cm, black},
    base_tensor/.style={draw=black, fill=white, line width=0.02cm, minimum size=0.6cm, inner sep=0.02cm},
    E/.style={base_tensor, rectangle, rounded corners=0.1cm},
    A/.style={base_tensor, circle},
    M/.style={base_tensor, rectangle, rounded corners=0.1cm, minimum width=1.2cm},
    B/.style={fill=myblue},
    Bd/.style={fill=myred},
    BBd/.style={fill=mypurple},
    H/.style={draw=mygreen, fill=mygreen, minimum size=1.0cm, text=black},
    proj/.style={
        draw=black, fill=white, line width=0.02cm,
        regular polygon, regular polygon sides=3, 
        inner sep=0pt, minimum size=0.45cm, rounded corners=0.1cm
    },
    P/.pic={
        \node[proj, shape border rotate=180] (P) at (0.5,-0.8) {\small $P$};
        \draw[thickline] (0,-0.4) to[out=270,in=110] (P.north west);
        \draw[line] (1,-0.4) to[out=270,in=70] (P.north east);
        \draw[thickline] ([yshift=0.1cm]P.south) -- +(0,-0.2);
        \node[proj, shape border rotate=180] (P) at (0.5,-0.8) {\small $P$};
    },
    Q/.pic={
        \node[proj] (Q) at (0.5,0.8) {\small $Q$};
        \draw[thickline] (0,0.4) to[out=90,in=250] (Q.south west);
        \draw[line] (1,0.4) to[out=90,in=290] (Q.south east);
        \draw[thickline] ([yshift=-0.1cm]Q.north) -- +(0,0.2);
        \node[proj] (Q) at (0.5,0.8) {\small $Q$};
    }
}

\usepackage[colorlinks=true,linkcolor=blue,citecolor=blue]{hyperref}
\usepackage{cleveref}

\crefname{subsection}{sec.}{secs.}
\crefname{figure}{fig.}{figs.}
\crefname{equation}{eq.}{eqs.}

\definecolor{mygreen}{RGB}{51,180,76}
\definecolor{myblue}{RGB}{128,200,255}

\begin{document}

\title{Optimized Tensor-Network Renormalization for Quantum Dynamics: Resolving the Spectral Function of $\mathrm{K_2Co(SeO_3)_2}$}

\author{Jiahang Hu}
\affiliation{Beijing National Laboratory for Condensed Matter Physics and Institute of Physics,
Chinese Academy of Sciences, Beijing 100190, China.}
\affiliation{School of Physical Sciences, University of Chinese Academy of Sciences,
Beijing 100049, China.}

\author{Runze Chi}
\affiliation{Division of Chemistry and Chemical Engineering, California Institute of Technology,
Pasadena, California 91125, USA}

\author{B. Normand}
\affiliation{PSI Center for Scientific Computing, Theory and Data, Paul Scherrer Institute, CH-5232 Villigen-PSI, Switzerland}

\author{Hai-Jun Liao}\email{navyphysics@iphy.ac.cn}
\affiliation{Beijing National Laboratory for Condensed Matter Physics and Institute of Physics,
Chinese Academy of Sciences, Beijing 100190, China.}

\author{T. Xiang}\email{txiang@iphy.ac.cn}
\affiliation{Beijing National Laboratory for Condensed Matter Physics and Institute of Physics,
Chinese Academy of Sciences, Beijing 100190, China.}
\affiliation{School of Physical Sciences, University of Chinese Academy of Sciences,
Beijing 100049, China.}

\begin{abstract}
Tensor-network methods have opened a powerful route for the study of dynamical spectral functions in two-dimensional quantum systems. However, existing approaches within the framework of infinite projected entangled-pair states construct the required renormalization tensors solely from the ground-state environment and can suffer from severe numerical instability. We identify the origin of this instability and introduce an excitation-tailored corner-transfer-matrix renormalization-group (ET-CTMRG) method to resolve it. By incorporating excitation tensors into the renormalization procedure, the method constructs a substantially more accurate effective Hamiltonian matrix and thereby yields reliable and well-converged excitation spectra. For Heisenberg antiferromagnets, it reduces truncation errors by orders of magnitude and for the particularly complex case of the supersolid phase in the triangular-lattice XXZ magnet $\mathrm{K_2Co(SeO_3)_2}$, it achieves excellent quantitative agreement with inelastic neutron-scattering measurements. ET-CTMRG therefore provides a robust framework for investigating the dynamical properties of strongly correlated quantum systems.
\end{abstract}

\maketitle

\textit{Introduction---}The excitation spectrum is central to understanding the dynamical response of quantum many-body systems and to interpreting spectroscopic experiments such as inelastic neutron scattering (INS) and resonant inelastic x-ray scattering (RIXS). Tensor-network approaches have emerged as a leading framework for computing dynamical spectra \cite{1995Ostlund, Haegeman2013, Vanderstraeten2015, Vanderstraeten2019,Liao2019, ponsioen2020, ponsioen2022, Tu2024, Arias_Espinoza2024}, which are especially valuable for highly frustrated magnetic and strongly interacting electronic systems~\cite{tri_afm_1, tri_afm_2, Zhang2024, kcoseo_1, tri_afm_3}. In particular, the methodology developed in Refs.~\cite{ponsioen2020, ponsioen2022, tri_afm_1, tri_afm_2}, which combines infinite Projected Entangled-Pair States (iPEPS), Corner-Transfer-Matrix Renormalization Group (CTMRG)~\cite{CTMRG_1,CTMRG_2,CTMRG_3}, and automatic differentiation (AD)~\cite{Liao2019}, has enabled the quantitative comparison of numerical spectral functions with INS data measured for a range of quasi-two-dimensional (2D) magnetic materials~\cite{tri_afm_1,tri_afm_2,tri_afm_3,kcoseo_1}. It has also enabled the reliable prediction of spin excitation spectra in emerging quantum phases, including model altermagnets~\cite{Liu2025} and quantum spin liquids~\cite{Tan2024,Wang2025,kagome_sl}. 

Despite these successes, existing tensor-network methods still suffer from intrinsic instabilities, which limit the numerical accuracy and can lead for some models to unphysical negative modes in the computed spectrum, signalling a methodological breakdown. A case in point is the triangular-lattice spin-supersolid material $\mathrm{K_2Co(SeO_3)_2}$~\cite{kcoseo_1,kcoseo_5,kcoseo_2,kcoseo_3,kcoseo_4}, which will serve as our testing ground below. We will show that these instabilities originate from the CTMRG construction of the effective Hamiltonian, in which the truncated edge and corner tensors are determined solely by the ground-state environment (to which we refer as GS-CTMRG)~\cite{ponsioen2020,ponsioen2022}. The resulting truncation cannot capture the excitation subspace in an optimal way. Although considering projector derivatives can improve the stability~\cite{second_derivative}, the required second-order derivatives and enlarged unit cells are computationally intensive and restricted to commensurate momenta, limiting their general applicability.

In this work, we introduce an excitation-tailored CTMRG (ET-CTMRG) method for calculating spectra. By guiding the truncation with both ground-state and excitation tensors, ET-CTMRG enables a much more accurate construction of the effective Hamiltonian matrix and thereby yields stable, reliable spectra across the full momentum space. The method avoids both unit-cell enlargement and second-order derivatives. Benchmark calculations for square- and triangular-lattice antiferromagnets demonstrate that ET-CTMRG significantly outperforms GS-CTMRG in both accuracy and numerical stability. For the highly anisotropic quantum magnet $\mathrm{K_2Co(SeO_3)_2}$, where the GS-CTMRG spectra are unstable, we apply ET-CTMRG to compute excitation spectra in quantitative agreement with INS experiments.

\begin{figure}[t]
\centering
\includegraphics[width=\columnwidth]{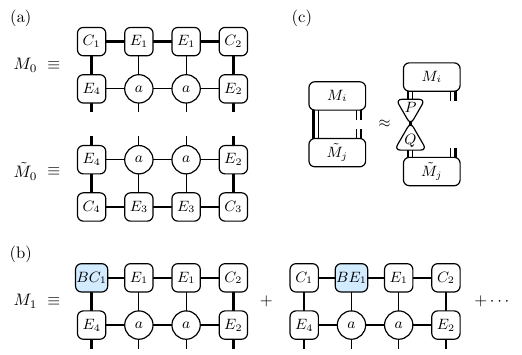}
\caption{Graphical representation of half-CTM tensors and projectors. The full 4$\times$4 CTM network is partitioned into upper and lower halves, each forming a 2$\times$4 network. (a) $M_0$ and $\tilde M_0$, defined solely from the ground-state tensors for the upper and lower half-CTM blocks, respectively. (b) $M_1$, defined as the sum of eight tensor networks, two of which are shown explicitly, with a single $B$ tensor placed at each of the eight possible positions. (c) Construction of the projectors $P$ and $Q$, which truncate all $M_i$ and $\tilde{M}_j$ simultaneously.}
\label{fig:configs}
\end{figure}

\textit{Methods---}In the iPEPS framework, a translationally invariant ground state is parameterized by a rank-5 tensor, $A$. Under the single-mode approximation, an excited state with momentum {\bf k} is then constructed as a superposition of local perturbations~\cite{Vanderstraeten2015},
\begin{equation}
    |\Phi_{k}(B)\rangle = \sum_{r} e^{i \mathbf{k} \cdot \mathbf{r}} |\Phi_{r}(B) \rangle,
\end{equation}
where $|\Phi_{r}(B)\rangle$ is formed by replacing one ground-state $A$ tensor, at location {\bf r}, with a local excitation tensor, $B$. Computing excitation spectra necessitates the evaluation of effective Hamiltonian and norm matrices
\begin{eqnarray}
\mathcal{H}^\text{eff}_{mn} & = & \langle \Phi_{k}(B^\dagger_m) |(H - E_0)| \Phi_{k}(B_n) \rangle , \\
\mathcal{N}^\text{eff}_{mn} & = & \langle \Phi_{k}(B^\dagger_m)| \Phi_{k}(B_n) \rangle ,
\end{eqnarray}
where $E_0$ is the ground-state energy and $B_n$ the $n$th basis tensor in the tangent space of the ground state~\cite{tangent_space}. The excitation energies, $E_p$, and associated eigenstates, $v_p$, are determined by solving the generalized eigenequation
\begin{equation}
    \mathcal{H}^\text{eff} v_p = E_p \mathcal{N}^\text{eff} v_{p}.
\label{eq:GEVD}
\end{equation}

The stability and reliability of the solutions to Eq.~\eqref{eq:GEVD} depend critically on the accuracy of the effective Hamiltonian and norm matrices. Evaluating these requires infinite summations over the 2D lattice, which is addressed in GS-CTMRG~\cite{ponsioen2020, ponsioen2022} by contracting the ground-state PEPS and summing over the excited states simultaneously, as summarized in Sec.~S1 of the Supplemental Material (SM) \cite{sm}. At each renormalization step, a pair of isometric operators, the projectors $P$ and $Q$, is introduced to truncate the basis space of the local tensors. These are determined by minimizing a cost function constructed solely within the ground-state manifold,
\begin{equation}
 \min_{\mathrm{rank}(PQ) \leq \chi} \Big| {M}_0^\dagger (I - PQ) \tilde{{M}}_0 \Big|,
\end{equation}
where $| \dots |$ denotes the Frobenius norm, $\chi$ is the CTM bond dimension, and $M_0$ and $\tilde{M}_0$ are the upper and lower halves of the full 4$\times$4 CTM network of the ground state [Fig.~\ref{fig:configs}(a)]. Although GS-CTMRG provides a straightforward scheme for evaluating $\mathcal{H}^\text{eff}$ and $\mathcal{N}^\text{eff}$, constructing $P$ and $Q$ solely from the ground-state PEPS is intrinsically inadequate for truncating tensor networks containing excitation tensors. This mismatch introduces systematic errors into both $\mathcal{H}^\text{eff}$ and $\mathcal{N}^\text{eff}$, which can lead to instabilities in solving Eq.~\eqref{eq:GEVD}.

We recognize that $P$ and $Q$ should be determined by incorporating the renormalization effects of excitation tensors explicitly in the contraction of excited states. To this end, we include all CTM configurations generated by the local excitation tensors $B$ and $B^\dagger$. For the upper half of the CTM network, these fall into three distinct classes. We define $M_1$ as the sum of the eight configurations in which a single $B$ tensor appears in one of the 2$\times$4 local tensors [Fig.~\ref{fig:configs}(b)], $M_2$ as the analogous sum containing a single $B^\dagger$, and $M_3$ as the sum over all configurations containing both $B$ and $B^\dagger$ within the upper-half CTM network. The lower-half CTM networks $(\tilde M_1,\tilde M_2,\tilde M_3)$ are defined analogously. Including these contributions together with the ground-state terms ($M_0$ and $\tilde M_0$), as represented in Fig.~\ref{fig:configs}(c), leads naturally to the generalized cost function for determining the projectors,
\begin{equation}
\min_{\mathrm{rank}(PQ) \leq \chi} \sum_{i,j=0}^{3} \Big| {M}_i^\dagger (I - PQ) \tilde{{M}}_j \Big|^2.
\end{equation}

It can be shown that this optimization is equivalent to minimizing the Frobenius error
\begin{equation}
 \varepsilon = \big| K^\dagger (I - PQ) \tilde K \big|,
\label{eq:Frobenius_error}
\end{equation}
where matrices $K$ and $\tilde K$ are defined through the relations $\sum_{i=0}^3 {M}_i {M}_i^\dagger = K K^\dagger$ and $\sum_{i=0}^3 \tilde{{M}}_i \tilde{{M}}_i^\dagger = \tilde K \tilde K^\dagger$. A derivation of Eq.~(\ref{eq:Frobenius_error}) is provided in Sec.~S2 of the SM \cite{sm}. By the Eckart--Young--Mirsky theorem~\cite{Eckart_Young_1936,Mirsky1960}, the optimal truncation projectors are obtained from the singular value decomposition of $K^{\dagger} \tilde K$, giving 
\begin{equation}
\quad P = \tilde K V \Lambda^{-1/2}, \qquad Q = \Lambda^{-1/2} U^\dagger K^\dagger,
\end{equation}
where $(\Lambda,U,V)$ denote the truncated (largest $\chi$) singular values and the corresponding singular vectors.

These steps constitute the core of the ET-CTMRG algorithm. Relative to GS-CTMRG, the only modification is replacing $M_0$ with $K$. This change is physically crucial, in that $K$ encodes the combined weight of all relevant excitation configurations in a compact way, tailoring the truncation projectors to preserve the excited-state subspace. $K$ then serves as the key entity transferring excitation information into the CTMRG coarse-graining. Because this replacement preserves the overall structure of the optimization, the computational cost remains very similar to that of GS-CTMRG.

Nevertheless, ET-CTMRG does not share all the attributes of GS-CTMRG. We have recently introduced a method achieving a dramatic speed-up of GS-CTMRG spectral calculations by avoiding the recomputation of $P$ and $Q$ \cite{kagome_sl}, and applied it to achieve a qualitative advance in understanding the spectrum of the kagome Heisenberg antiferromagnet (KHAF). This approach cannot be applied to ET-CTMRG, which requires new excitation-tailored projectors at each step, and this leads to a strategy call. If the GS-CTMRG spectrum of a model is adequately stable, in the sense we benchmark next, then the accelerated approach can be applied, but if it suffers intrinsic instability then only ET-CTMRG is applicable. 

\begin{figure}[t]
\centering
\includegraphics[width=0.76\columnwidth]{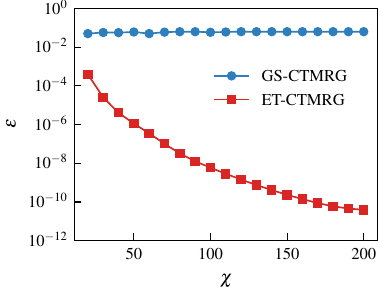}
 \caption{Frobenius error, $\varepsilon$, shown as a function of the CTM bond dimension, $\chi$, obtained with GS-CTMRG and ET-CTMRG for the square-lattice Heisenberg antiferromagnet (SLHAF) at momentum $(0.2\pi,0.3\pi)$ and $D = 3$.}
\label{fig:square_chi_convergence}
\end{figure}

\textit{Benchmark and application---}We first benchmark ET-CTMRG against GS-CTMRG for a very well known model, the spin-$1/2$ Heisenberg antiferromagnet (HAF) on the square lattice (SL). The following illustration is valid for any general momentum point, as well as for each accessible iPEPS bond dimension, and so we choose the values ${\bf k} = (0.2\pi,0.3\pi)$ and $D = 3$. Figure \ref{fig:square_chi_convergence} compares the Frobenius error, $\varepsilon$, as a function of $\chi$ for the two methods. The difference is remarkable. For GS-CTMRG, $\varepsilon$ remains large and shows little improvement with increasing $\chi$, indicating that the truncation does not converge systematically toward the relevant excitation subspace. By contrast, ET-CTMRG exhibits clear and systematic convergence as $\chi$ increases, with the error reduced to approximately $10^{-11}$ at $\chi = 200$, corresponding to an improvement by some ten orders of magnitude.

\begin{figure}[t]
\centering
\includegraphics[width=\columnwidth]{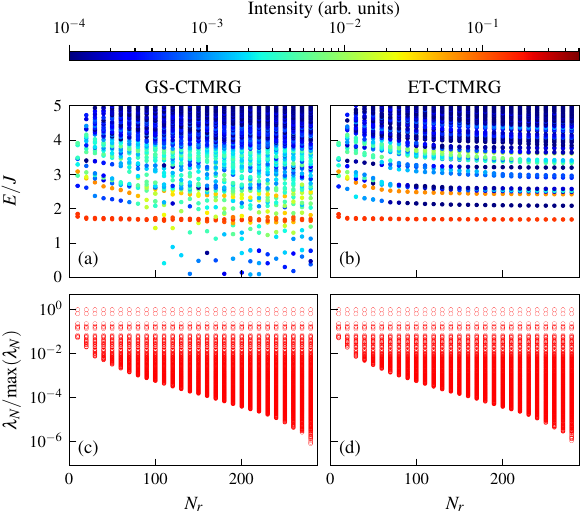}
 \caption{Comparison of excitation spectra (a,b) and normalized norm-matrix eigenvalues (c,d) as functions of the number of retained states, $N_r$, obtained with GS-CTMRG (a,c) and ET-CTMRG (b,d) for the SLHAF, computed at momentum $(0.2\pi,0.3\pi)$ with $D = 3$ and $\chi = 60$. In panels (a) and (b), the color indicates the spectral weight in each eigenstate.}
\label{fig:square_basis_convergence}
\end{figure}

Next we introduce a parameter that is mentioned less often when calculating spin excitations by iPEPS, namely the number of states retained to represent the spectral function. The retained states are selected according to the eigenvalues of the norm matrix. $N_r$ should be large for an accurate image, but retaining many states risks that most have very low spectral weights with concomitant uncertainties in energy. Most work in the field to date has used smaller $N_r$ values for an accurate representation of the high-weight states. Figures \ref{fig:square_basis_convergence}(a) and \ref{fig:square_basis_convergence}(b) compare the excitation spectra of the SLHAF obtained with GS- and ET-CTMRG as $N_r$ increases, working again at momentum $(0.2\pi,0.3\pi)$ and $D = 3$ with $\chi = 60$. In another striking contrast, the excitation energies within  GS-CTMRG become widely scattered beyond $N_r = 100$ and show no sign of convergence as more states are retained. In contrast, the ET-CTMRG spectrum stabilizes systematically and continues to converge as the excited-state subspace is enlarged. 

The instabilities encountered in GS-CTMRG are often attributed to an ill-conditioned norm matrix, meaning that the numerical errors cause the eigenvalues of $\mathcal{N}^{\mathrm{eff}}$ to become very small or negative~\cite{ponsioen2020,ponsioen2022}. This has motivated the common practice of further truncating the excitation basis with intent to regularize the generalized eigenvalue problem by discarding such states. However, Figs.~\ref{fig:square_basis_convergence}(c) and \ref{fig:square_basis_convergence}(d) show that the norm-matrix eigenvalues obtained with GS- and ET-CTMRG are nearly identical, and not particularly small. Clearly, the severe instability of GS-CTMRG in our example originates rather from numerical errors in $\mathcal{H}^\text{eff}$, which must be computed with an accuracy comparable to $\mathcal{N}^{\mathrm{eff}}$ to avoid its eigenvalues remaining scattered or even becoming unphysical as $N_r$ is varied. We deduce that the ET-CTMRG approach of incorporating excitation information directly into the construction of $P$ and $Q$ succeeds because its faithful renormalization of the excited-state manifold produces a much more accurate $\mathcal{H}^{\mathrm{eff}}$. The resulting smoothly convergent and physically meaningful excitation spectrum [Fig.~\ref{fig:square_basis_convergence}(b)] highlights the importance of treating the ground- and excited-state manifolds on an equal footing during the coarse-graining procedure, and this is reinforced by the further benchmarks shown in Sec.~S2 of the SM \cite{sm}.

\begin{figure}[t]
\centering
\includegraphics[width=\columnwidth]{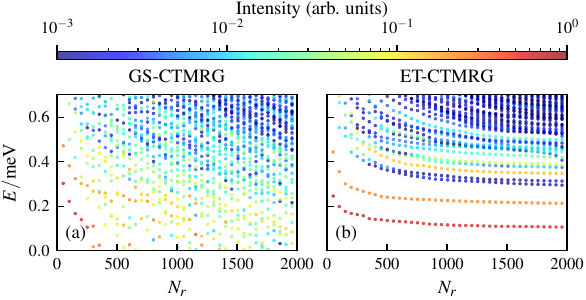}
 \caption{Comparison of the excitation spectra computed for the triangular-lattice XXZ model of $\mathrm{K_2Co(SeO_3)_2}$ at the M point and zero field using GS-CTMRG (a) and ET-CTMRG (b), shown as functions of $N_r$ for $D = 4$ and $\chi = 100$.}
\label{fig:KSCO_basis_convergence}
\end{figure}

The remarkable difference between GS- and ET-CTMRG for an isotropic and unfrustrated model such as the SLHAF begs the question of which physical ingredients drive the instability of the effective Hamiltonian matrix. Obvious candidates would include frustration, anisotropy, and special momentum points. In Sec.~S3 of the SM \cite{sm} we show benchmarks similar to Fig.~\ref{fig:square_basis_convergence} for two paradigm frustrated models, the triangular-lattice (TL) and kagome HAFs. At the M points of both models, the GS-CTMRG spectra are at least as stable as for the SLHAF, retaining their form out to $N_r \approx 1000$ for $D = 4$. These results indicate that frustration is not a leading factor in driving the instability and validate earlier iPEPS analyses of the spectra of both models, including our own \cite{tri_afm_1,kagome_sl}. By contrast, the highly anisotropic TL material $\mathrm{K_2Co(SeO_3)_2}$~\cite{kcoseo_1, kcoseo_2, kcoseo_3,kcoseo_4,kcoseo_5}, which is described by a strongly easy-axis XXZ model, presents a case study with an extremely unstable GS-CTMRG spectrum. 

$\mathrm{K_2Co(SeO_3)_2}$ has attracted attention because the noncollinear Y and V phases of its field-induced phase diagram realize the spin-supersolid state \cite{Wessel2005,Melko2005,Heidarian2005, Boninsegni2005,Heidarian2010,Sellmann2015}, which has simultaneous diagonal and off-diagonal magnetic order (respectively of the longitudinal and transverse components). $\mathrm{K_2Co(SeO_3)_2}$ has a dominant energy scale of $J_{z} = 3.1$ meV and the very strongly anisotropic interaction ratio $J_{xy}/J_{z} = 0.07$ that places it close to the Ising limit, where one might assume from the macroscopic classical degeneracy of the ground manifold \cite{Wannier1950,Houtappel1950} that the excitation spectrum of the quantum model is exquisitely sensitive to truncation errors. A previous GS-CTMRG calculation~\cite{kcoseo_1} found highly unstable spectra with no convergence tendency, retaining only 50 excited states for the $D = 3$ iPEPS (approximately one tenth of the full excitation basis) to suppress the instability. This truncation not only prevented a quantitative description of the measured spectra~\cite{kcoseo_2,kcoseo_3,kcoseo_4} but also called the qualitative predictions into question, inspiring further recent analysis by a number of approaches \cite{kcoseo_6,kcoseo_7,kcoseo_8,kcoseo_9}.

\begin{figure*}[t]
\centering
\includegraphics[width=\textwidth]{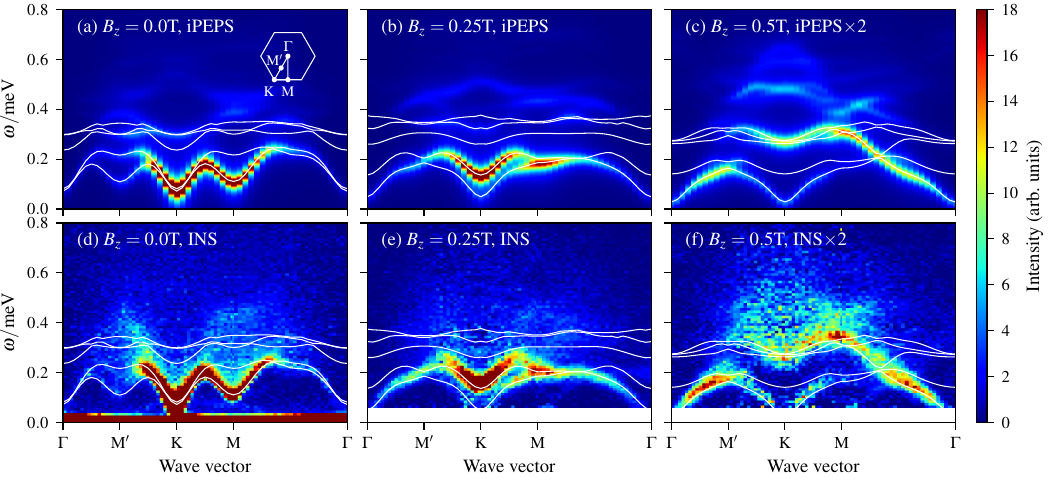}
\caption{Spin excitation spectra of $\mathrm{K_2Co(SeO_3)_2}$ computed using ET-CTMRG with $D = 4$ and $\chi = 100$ and shown along the high-symmetry momentum path $\Gamma$KM$\Gamma$ [inset panel (a)]. Panels (a–c) display the computed INS intensity at applied fields equivalent to $B_z = 0$, 0.25, and 0.5~T, respectively, and panels (d–f) the experimental data of Ref.~\cite{kcoseo_3}. Measurements at $B_z = 0$~T were performed on the spectrometer AMATERAS (J-PARC, Japan) and at finite fields on LET (ISIS, UK); the intensity at $B_z = 0.5$~T is multiplied by two to match the experimental presentation.}
\label{fig:KCoSeO_spectra}
\end{figure*}

In Fig.~\ref{fig:KSCO_basis_convergence} we illustrate both the dramatic instability of the GS-CTMRG framework when applied to the $\mathrm{K_2Co(SeO_3)_2}$ model and the complete contrast with the ET-CTMRG spectrum, which remains completely stable and approaches a well-converged limit as $N_r$ is increased. It is clear that excitation-tailoring is not merely a technical refinement, but constitutes a necessary ingredient for obtaining quantitatively reliable excitation spectra in certain classes of quantum magnet.

We complete our study by using ET-CTMRG to compute the full excitation spectrum of $\mathrm{K_2Co(SeO_3)_2}$ in the spin-supersolid Y phase for comparison with INS, which measures the quantity~\cite{Squires2012} 
\begin{equation}
I(\mathbf{k},\omega) \propto \tfrac{1}{2} g_{ab}^2 (S^{xx} + S^{yy}) (\mathbf{k},\omega) + g_c^2 S^{zz}(\mathbf{k},\omega).
\end{equation}
The strong easy-axis anisotropy of the spin interactions in $\mathrm{K_2Co(SeO_3)_2}$ is accompanied by a highly anisotropic $g$-tensor ($g_c/g_{ab} \approx 4$~\cite{kcoseo_3,kcoseo_4}), which means that the INS response is dominated by the longitudinal channel. Figures \ref{fig:KCoSeO_spectra}(a-c) present our calculation of $I(\mathbf{k},\omega)$ at three magnetic fields in the Y phase, $B_z = 0$, 0.25, and 0.5~T, and Figs.~\ref{fig:KCoSeO_spectra}(d-f) the corresponding INS measurements provided by the authors of Ref.~\cite{kcoseo_3}. In Sec.~S4 of the SM \cite{sm} we use our iPEPS calculations to separate the transverse and longitudinal components, which makes it easier to visualize their contributions to the unweighted dynamical structure factor, $S^{\mathrm{tot}} (\mathbf{k},\omega) = S^{xx}(\mathbf{k},\omega) + S^{yy}(\mathbf{k},\omega) + S^{zz}(\mathbf{k},\omega)$.

At zero field, the iPEPS spectrum exhibits two low-energy magnon branches [Fig.~\ref{fig:KCoSeO_spectra}(a)], of which only the lower is resolved experimentally [Fig.~\ref{fig:KCoSeO_spectra}(d)], and our results reproduce both its dispersion and spectral weight with quantitative accuracy. We remark here that iPEPS cannot capture the gapless nature of the Goldstone mode at the $\Gamma$ and K points due to the finite value of $D$, and we extrapolate the energies of this mode at the high-symmetry points to $D = 5$ in Sec.~S4 of the SM \cite{sm}. Both at M and at the symmetry-equivalent M$^\prime$ point mid-way between $\Gamma$ and K, the two branches are well separated and both develop the minima often referred to as ``rotons'' despite their commensurate momenta. At M, the lower branch (deeper minimum) originates from longitudinal fluctuations ($S^{zz}$) and the upper from transverse ones, explaining its near-invisibility to INS. Neither the strong splitting of these modes nor their minima are found in linear spin-wave theory. The iPEPS spectrum shows one distinct set of higher modes, which can be associated with rather well-defined two-magnon excitations, before changing its form to broader scattering continua. 

As the field is increased, the longitudinal branch moves up in energy, becoming gapped at K, and the minima at M and M$^\prime$ disappear [Figs.~\ref{fig:KCoSeO_spectra}(b,e)]. The transverse branch remains almost unchanged, forming the Goldstone mode, but has very little INS intensity. At $B_z = 0.5$~T, approaching the 0.8~T transition from the Y to the UUD phase, the longitudinal mode loses spectral weight and mixes with the two-magnon excitations, which in turn merge with the considerably broadened scattering continuum [Figs.~\ref{fig:KCoSeO_spectra}(c,f)]. Further comments on the physics of these spectra may be found in Sec.~S4 of the SM \cite{sm}. The excellent quantitative agreement between the measured and calculated spectra demonstrates that ET-CTMRG is capable of capturing both the sharply defined excitations and the continuum scattering features in this strongly anisotropic quantum magnet.

\textit{Conclusion---}We have introduced the ET-CTMRG method for computing excitation spectra within the iPEPS framework. By incorporating excitation tensors directly into the construction of the CTMRG truncation projectors, ET-CTMRG ensures that the renormalized environment captures the relevant excited-state manifold in a faithful manner. At the same time, it preserves the computational cost and implementation simplicity of GS-CTMRG while markedly improving the accuracy of the effective Hamiltonian and eliminating the numerical instabilities inherent to the ground-state-based truncation scheme, which can become very severe in some systems.

Benchmark calculations for Heisenberg models show that ET-CTMRG reduces the truncation error systematically and yields well-converged spectra as the excited-state basis is enlarged. Our studies demonstrate that the primary source of instability is not the ill-conditioning of the norm matrix, but insufficient accuracy of the effective Hamiltonian matrix. For the triangular-lattice material $\mathrm{K_2Co(SeO_3)_2}$, where GS-CTMRG fails to produce converged spectra, ET-CTMRG yields spectra in quantitative agreement with INS measurements. These results establish ET-CTMRG as a reliable and predictive framework for dynamical spectroscopy and highlight its potential as a key component of future tensor-network approaches to excitation spectra in strongly correlated quantum materials.
 
\bigskip
\textit{Acknowledgments.---}We thank G. Chen and A. Zheludev for valuable discussions. We are grateful to M. Zhu and A. Zheludev for providing their INS data for $\mathrm{K_2Co(SeO_3)_2}$. This work was supported by the Strategic Priority Research Program of the Chinese Academy of Sciences (CAS, under Grant No.~XDB0500202), the National Key Research and Development Project of China (Grants No.~2024YFA1408604, 2021ZD0301800, and 2022YFA1403900), the National Natural Science Foundation of China (Grants No.~12488201, 12322403, and 12347107), the CAS Project for Young Scientists in Basic Research (Grant No.~YSBR-150), and the Youth Innovation Promotion Association CAS (Grant No.~2021004). The numerical calculations in this work were carried out on the ORISE Supercomputer of the CAS.

\bibliography{ipeps_adv}

\setcounter{figure}{0}
\renewcommand{\thefigure}{S\arabic{figure}}
\setcounter{section}{0}
\renewcommand{\thesection}{S\arabic{section}}
\setcounter{equation}{0}
\renewcommand{\theequation}{S\arabic{equation}}
\setcounter{table}{0}
\renewcommand{\thetable}{S\arabic{table}}

\onecolumngrid
\vskip 40mm

\noindent
{\bf{\large{Supplemental Material to accompany the manuscript}}}

\vskip 4mm

\noindent
{\bf{\large{Optimized Tensor-Network Renormalization for Quantum Dynamics: Resolving the Spectral Function of $\mathrm{K_2Co(SeO_3)_2}$}}}

\vskip 6mm

\noindent
Jiahang Hu, Runze Chi, B. Normand, Hai-Jun Liao, and T. Xiang

\vskip 10mm

\twocolumngrid

\section{Calculating Excitation Spectra by CTMRG: An Overview}
\label{secs1}

\subsection{Ground-State Calculation}

The infinite Projected Entangled-Pair State (iPEPS) is a tensor-network Ansatz designed to represent two-dimensional (2D) quantum states directly in the thermodynamic limit. The square-lattice Heisenberg antiferromagnet (SLHAF) studied in the main text has the spin Hamiltonian 
\begin{equation}
H = J \sum_{\langle i j \rangle} \vec{S}_i \cdot \vec{S}_j ,
\end{equation}
where $\vec{S}_i$ denotes the spin operator at site $i$, $J > 0$ is the spin-spin interaction, and $\langle ij \rangle$ denotes a sum over only adjacent sites $i$ and $j$ for the nearest-neighbor variant of the model. The ground state is a two-sublattice N\'eel-ordered state, for which we use a two-sublattice iPEPS parametrized by the two tensors $A_1$ and $A_2$,
\begin{equation}
  \begin{tikzpicture}[scale=0.8, transform shape]
    
    \begin{scope}
        \node[anchor=east, font=\large] at (-1.5, -1.5) {$\ket{\Psi_0} =$};
        
        \foreach \i in {0,...,3} {
            \draw[line] (\i, 0.5) -- (\i, -3.5);
            \draw[line] (-0.5, -\i) -- (3.5, -\i);
        }

        \foreach \x in {0,...,3} {
            \foreach \y in {0,...,3} {
                \pgfmathsetmacro{\sublattice}{int(mod(\x+\y, 2) == 0 ? 1 : 2)}
                \draw[line] (\x, -\y) -- (\x+0.4, -\y-0.3);
                \node[A] at (\x, -\y) {$A_{\sublattice}$};
            }
        }

        \node at (4, -1.5) {$\dots$};
        \node at (-1, -1.5) {$\dots$};
        \node at (1.5, 1.0) {$\vdots$};
        \node at (1.5, -4.0) {$\vdots$};
    \end{scope}

\end{tikzpicture} \nonumber
\end{equation}
The triangular-lattice Heisenberg antiferromagnet (TLHAF) and the kagome-lattice Heisenberg antiferromagnet (KHAF) are obtained by placing the same Hamiltonian respectively on triangular and kagome geometries, and here we will again discuss only the nearest-neighbor versions of both models. 

For the material $\mathrm{K_2Co(SeO_3)_2}$, the physics is captured by an XXZ model on the triangular lattice,
\begin{equation}
H = \sum_{\langle i j \rangle} \big[ J_{xy} \left( S_{i}^{x} S_{j}^{x} + S_{i}^{y} S_{j}^{y} \right) + J_{z} S_{i}^{z} S_{j}^{z} \big] - h \sum_i S_{i}^{z}, \nonumber
\label{Ham}
\end{equation}
where $h = g_c \mu_B B$, $B$ is the external magnetic field applied along the crystalline $c$ axis (i.e.~perpendicular to the TL plane), and $\langle i j \rangle$ denotes again only pairs of nearest-neighbor sites. In our calculation, we adopt the parameters determined in experiment \cite{kcoseo_3} as $J_{z} = 3.1$ meV, $J_{xy} = 0.217$ meV, and the $g$-factor $g_c = 7.9$. The ground state of this model exhibits three-sublattice order at all fields below saturation, and hence we use an iPEPS in which three sites, one from each sublattice, are grouped into a single effective site, 
\begin{equation}
  \begin{tikzpicture}[scale=0.8, transform shape]

    \begin{scope}[scale=0.8, transform shape]

        \clip (-0.6,-3.6) rectangle (3.5,0.7);
        
        \foreach \i in {0,...,4}
            \draw[line, dashed, line width=0.01cm, gray] (-1, {-\i*sqrt(3)/2}) -- (6, {-\i*sqrt(3)/2});

        \foreach \i in {-2,...,6}
            \draw[line, dashed, line width=0.01cm, gray] ({\i+4}, {4*sqrt(3)}) -- ({\i-8}, {-8*sqrt(3)});

        \foreach \i in {-2,...,6}
            \draw[line, dashed, line width=0.01cm, gray] ({\i+4}, {-4*sqrt(3)}) -- ({\i-8}, {8*sqrt(3)});

        \foreach \i in {0,...,3}
            \foreach \j in {0,...,3}
                \node[draw=myblue, fill=myblue, line width=0cm, regular polygon, regular polygon sides=3, minimum size=1.15cm] at ({\i*3/2}, {\i*sqrt(3)/2-\j*sqrt(3)-sqrt(3)/3}) {};

        \foreach \i in {0,...,3}
            \draw[line] ({\i*3/2}, -4) -- ({\i*3/2}, 1);

        \foreach \i in {0,...,3}
            \draw[line] (-1, {-1*sqrt(3)/3-\i*sqrt(3)-sqrt(3)/3}) -- (6, {6*sqrt(3)/3-\i*sqrt(3)-sqrt(3)/3});

        \foreach \i in {0,...,3}{
            \foreach \j in {0,...,3}{
                \draw[line] (\i*3/2, {\i*sqrt(3)/2-\j*sqrt(3)-sqrt(3)/3}) -- (\i*3/2+0.4, {\i*sqrt(3)/2-\j*sqrt(3)-sqrt(3)/3-0.3});
                \node[A] at ({\i*3/2}, {\i*sqrt(3)/2-\j*sqrt(3)-sqrt(3)/3}) {$A$};
            }
        }

    \end{scope}

    \begin{scope}

        \node[anchor=east, font=\large] at (-1.5*0.8, -1.45*0.8) {$\ket{\Psi_0} =$};
    
        \node at (4.0*0.8, -1.45*0.8) {$\dots$};
        \node at (-1.1*0.8, -1.45*0.8) {$\dots$};
        \node at (1.45*0.8, 1.2*0.8) {$\vdots$};
        \node at (1.45*0.8, -4.0*0.8) {$\vdots$};

    \end{scope}

    \begin{scope}[shift={(5.2,-1.45*0.8+1)}]

        \node[] at (-1.5,-1) {$=$};
        
        \foreach \i in {0,1,2} {
            \draw[line] (\i, 0.5) -- (\i, -2.5);
            \draw[line] (-0.5, -\i) -- (2.5, -\i);
        }

        \foreach \x in {0,1,2} {
            \foreach \y in {0,1,2} {
                \draw[line] (\x, -\y) -- (\x+0.4, -\y-0.3);
                \node[A] at (\x, -\y) {$A$};
            }
        }

        \node at (3, -1) {$\dots$};
        \node at (-1, -1) {$\dots$};
        \node at (1, 1.0) {$\vdots$};
        \node at (1, -3.0) {$\vdots$};
    \end{scope}

\end{tikzpicture} \nonumber
\end{equation}
Although this Ansatz has only one local tensor ($A$) per unit cell, the physical index of each tensor has dimension $d = 8$ to represent the degrees of freedom of the three grouped $S = 1/2$ spins. In previous work~\cite{tri_afm_1,tri_afm_2,tri_afm_3} we found that, for a given bond dimension $D$, this representation yields a ground-state energy slightly higher than that of a three-sublattice iPEPS with separate tensors; however, for calculating the excitation spectrum it has the advantage of being more stable, because the process of obtaining the effective excitation Hamiltonian requires fewer iterations (as we detail in Sec.~\ref{secs1}B).

Once the iPEPS Ansatz for the ground state is chosen, expectation values are computed using the corner-transfer-matrix (CTM) renormalization-group (RG) method~\cite{CTMRG_1, CTMRG_2, CTMRG_3} and the variational parameters (the local tensors) are optimized by gradient-based methods such as automatic differentiation (AD)~\cite{Liao2019}. For simplicity of presentation, we illustrate how to use the CTMRG method to calculate the norm of a translationally invariant, single-sublattice iPEPS, whose wave function is
\begin{equation}
  \begin{tikzpicture}[scale=0.8, transform shape]
    
    \begin{scope}
        \node[anchor=east, font=\large] at (-1.5, -1) {$\ket{\Psi_0} =$};
        
        \foreach \i in {0,1,2} {
            \draw[line] (\i, 0.5) -- (\i, -2.5);
            \draw[line] (-0.5, -\i) -- (2.5, -\i);
        }

        \foreach \x in {0,1,2} {
            \foreach \y in {0,1,2} {
                \draw[line] (\x, -\y) -- (\x+0.4, -\y-0.3);
                \node[A] at (\x, -\y) {$A$};
            }
        }

        \node at (3, -1) {$\dots$};
        \node at (-1, -1) {$\dots$};
        \node at (1, 1.0) {$\vdots$};
        \node at (1, -3.0) {$\vdots$};
    \end{scope}

\end{tikzpicture},  \nonumber
\end{equation}
where the rank-5 tensor $A$ has four virtual indices of bond dimension $D$ and one physical index of dimension $d$. The norm of the ground state can be computed by contracting
the infinite 2D double-layer tensor network
\begin{equation}
  \begin{tikzpicture}[scale=0.8, transform shape]
    
    \begin{scope}
        \node[anchor=east] at (-1.5, -1) {$\braket{\Psi_0} =$};
        
        \foreach \i in {0,1,2} {
            \draw[line] (\i, 0.5) -- (\i, -2.5);
            \draw[line] (-0.5, -\i) -- (2.5, -\i);
        }

        \foreach \x in {0,1,2} {
            \foreach \y in {0,1,2} {
                \node[A] at (\x, -\y) {$a$};
            }
        }

        \node at (3, -1) {$\dots$};
        \node at (-1, -1) {$\dots$};
        \node at (1, 1.0) {$\vdots$};
        \node at (1, -3.0) {$\vdots$};
    \end{scope}

\end{tikzpicture}, \nonumber
\end{equation}
where 
\begin{equation}
  \begin{tikzpicture}[scale=0.8, transform shape]

  \begin{scope}
    \draw[line] (-0.5,0) -- (0.5,0);
    \draw[line] (0,-0.5) -- (0,0.5);
    \node[A] at (0,0) {\small $a$};
    
    \node at (1.0,0) {\large $=$};
  \end{scope}

  \begin{scope}[shift={(2.0, 0.5)}]
    \draw[line] (0,0) -- (0,-1);
    
    \foreach \y in {0, -1} {
        \draw[line] (-0.5,\y) -- (0.5,\y);
        \draw[line] (-0.5,\y-0.3) -- (0.5,\y+0.3);
    }
    
    \node[A] at (0,0) {\small $A$};
    \node[A] at (0,-1) {\small $A^\dagger$};
  \end{scope}

\end{tikzpicture}  \nonumber
\end{equation}
is shorthand for the double-layer tensor $AA^\dagger$. This contraction cannot be performed exactly, and normally one employs CTMRG~\cite{CTMRG_1, CTMRG_2, CTMRG_3} to approximate the contraction of the infinite tensor network. The core concept is to find a set of boundary tensors, $\{C_i, E_i\}$ $(i = 1, \dots, 4)$, that represent the environment of the double-layer network,
\begin{equation}
  \begin{tikzpicture}[scale=0.8, transform shape]
    
    \begin{scope}
        
        \foreach \i in {0,1,2} {
            \draw[line] (\i, 0.5) -- (\i, -2.5);
            \draw[line] (-0.5, -\i) -- (2.5, -\i);
        }

        \foreach \x in {0,1,2} {
            \foreach \y in {0,1,2} {
                \node[A] at (\x, -\y) {$a$};
            }
        }

        \node at (3, -1) {$\dots$};
        \node at (-1, -1) {$\dots$};
        \node at (1, 1.0) {$\vdots$};
        \node at (1, -3.0) {$\vdots$};
        \node at (3.6, -1) {$\approx$};
    \end{scope}

    \begin{scope}[shift={(4.5,0)}]
        \foreach \i in {0,2} {
            \draw[thickline] (\i, 0) -- (\i, -2);
            \draw[thickline] (0, -\i) -- (2, -\i);
        }

        \draw[line] (1, 0) -- (1, -2);
        \draw[line] (0, -1) -- (2, -1);

        \node[E] at (0, 0) {$C_1$};
        \node[E] at (2, 0) {$C_2$};
        \node[E] at (2, -2) {$C_3$};
        \node[E] at (0, -2) {$C_4$};
        
        \node[E] at (1, 0) {$E_1$};
        \node[E] at (2, -1) {$E_2$};
        \node[E] at (1, -2) {$E_3$};
        \node[E] at (0, -1) {$E_4$};
        
        \node[A] at (1, -1) {$a$};

        \node[] at (0.5, 0.4) {$\chi$};
    \end{scope}

\end{tikzpicture}.
 \label{ctm_env}
\end{equation}

These boundary tensors have boundary bond dimension $\chi$, and to obtain them one performs an iterative RG process, where the $a$ tensors are absorbed into the boundary tensors at each step. As one example, in a left-move RG step, the update of $E_4$ is
\begin{equation}
  \begin{tikzpicture}[scale=0.8, transform shape]

  \begin{scope}
    \draw[line] (0,0) -- (0.7,0);
    \draw[thickline] (0,-0.7) -- (0,0.7);
    \node[E] at (0,0) {\small $E_4'$};
    \node at (1.25,0) {$=$};
  \end{scope}

  \begin{scope}[shift={(2.25,0)}]
    \draw[line] (0,0) -- (1,0);
    \draw[thickline] (0,0.4) -- (0,-0.4);

    \draw[line] (1,0) -- (1,0.4);
    \draw[line] (1,0) -- (1.7,0);
    \draw[line] (1,0) -- (1,-0.4);

    \node[E] at (0,0) {\small $E_4$};
    \node[A] at (1,0) {\small $a$};

    \pic at (0,0) {P};
    \pic at (0,0) {Q};
  \end{scope}

\end{tikzpicture}.  \nonumber
\end{equation}
In this step, the projectors $P$ and $Q$ truncate the new boundary dimension from $\chi D^2$ to $\chi$ in order to prevent its infinite growth. Correspondingly, $C_1$ is updated by absorbing an $E_1$ tensor and is truncated by $P$,
\begin{equation}
  \begin{tikzpicture}[scale=0.8, transform shape]

  \begin{scope}
    \draw[thickline] (0,0) -- (0,-0.7);
    \draw[thickline] (0,0) -- (0.7,0);
    \node[E] at (0,0) {\small $C_1'$};
  \end{scope}

  \begin{scope}[shift={(1.25,0)}]
    \node at (0,0) {$=$};
  \end{scope}

  \begin{scope}[shift={(2.25,0)}]
    \draw[thickline] (0,0) -- (0,-0.4);
    \draw[thickline] (0,0) -- (1.5,0);
    \draw[line] (1,0) -- (1,-0.4);
    
    \node[E] at (0,0) {\small $C_1$};
    \node[E] at (1,0) {\small $E_1$};

    \pic at (0,0) {P};
  \end{scope}

\end{tikzpicture}, \nonumber
\end{equation}
while the analogous update of $C_4$ involves $Q$,
\begin{equation}
  \begin{tikzpicture}[scale=0.8, transform shape]

  \begin{scope}
    \draw[thickline] (0,0) -- (0,0.7);
    \draw[thickline] (0,0) -- (0.7,0);
    \node[E] at (0,0) {\small $C_4'$};
  \end{scope}

  \begin{scope}[shift={(1.25,0)}]
    \node at (0,0) {$=$};
  \end{scope}

  \begin{scope}[shift={(2.25,0)}]
    \draw[thickline] (0,0) -- (0,0.4);
    \draw[thickline] (0,0) -- (1.5,0);
    \draw[line] (1,0) -- (1,0.4);
    
    \node[E] at (0,0) {\small $C_4$};
    \node[E] at (1,0) {\small $E_3$};

    \pic at (0,0) {Q};
  \end{scope}

\end{tikzpicture}. \nonumber
\end{equation}

To preserve as much relevant information as possible during the iterative RG process, the most important step in CTMRG is to construct the projectors $P$ and $Q$ that minimize the truncation error at each iteration. For general, non-symmetric tensor networks, a widely adopted approach is the algorithm developed by Corboz \textit{et al.}~\cite{CTMRG_3}. The construction of $P$ and $Q$ proceeds by first defining the upper and lower half-blocks, $M_0$ and $\tilde{M}_0$,
\begin{equation}
  \begin{tikzpicture}[scale=0.8, transform shape]

  \begin{scope}

    \foreach \i in {0,3} {
        \draw[thickline] (\i,0) -- (\i,-1.5);
    }

    \foreach \i in {1,2} {
        \draw[line] (\i,0) -- (\i,-1.5);
    }

    \draw[thickline] (0,0) -- (3,0);
    \draw[line] (0,-1) -- (3,-1);
    
    \node[E] at (0,0) {\small $C_1$};
    \node[E] at (3,0) {\small $C_2$};
    \node[E] at (1,0) {\small $E_1$};
    \node[E] at (2,0) {\small $E_1$};
    \node[E] at (3,-1) {\small $E_2$};
    \node[E] at (0,-1) {\small $E_4$};
    
    \foreach \x in {1,2}
        \node[A] at (\x,-1) {\small $a$};

    % \draw[thick, red, rounded corners] (-0.5,0.5) rectangle (3.5,-1.5);

    \foreach \i in {-2.5,-1.5} {
        \draw[thickline] (\i,-0.5) -- (\i,-1.0);
    }

    \foreach \i in {-2.4,-1.6} {
        \draw[line] (\i,-0.5) -- (\i,-1.0);
    }
    
    \node[M] at (-2.0,-0.5) {\small $M_0$};
    
    \node[font=\large] at (-0.9,-0.5) {$\equiv$};

  \end{scope}
  
  \begin{scope}[shift={(0,-0.5)}]

    \foreach \i in {0,3} {
        \draw[thickline] (\i,-1.5) -- (\i,-3);
    }

    \foreach \i in {1,2} {
        \draw[line] (\i,-1.5) -- (\i,-3);
    }

    \draw[line] (0,-2) -- (3,-2);
    \draw[thickline] (0,-3) -- (3,-3);

    \node[E] at (3,-3) {\small $C_3$};
    \node[E] at (0,-3) {\small $C_4$};
    \node[E] at (3,-2) {\small $E_2$};
    \node[E] at (1,-3) {\small $E_3$};
    \node[E] at (2,-3) {\small $E_3$};
    \node[E] at (0,-2) {\small $E_4$};
    
    \foreach \x in {1,2}
        \foreach \y in {2}
            \node[A] at (\x,-\y) {\small $a$};

    % \draw[thick, red, rounded corners] (-0.5,-1.5) rectangle (3.5,-3.5);

    \foreach \i in {-2.5,-1.5} {
        \draw[thickline] (\i,-2.0) -- (\i,-2.5);
    }

    \foreach \i in {-2.4,-1.6} {
        \draw[line] (\i,-2.0) -- (\i,-2.5);
    }
    
    \node[M] at (-2.0,-2.5) {\small $\tilde M_0$};

    \node[font=\large] at (-0.9,-2.5) {$\equiv$};
            
  \end{scope}

\end{tikzpicture} \, . \nonumber
\end{equation}
Then, one seeks a pair $P$ and $Q$ to truncate $M_0^\dagger \tilde M_0$, 
\begin{equation}
    \begin{tikzpicture}[scale=0.8, transform shape]

  \begin{scope}

    \draw[thickline] (-0.5,0.0) -- (-0.5,-1.2);
    \draw[line] (-0.4,0.0) -- (-0.4,-1.2);
    
    \draw[line] (0.4,0.0) -- (0.4,-0.5);
    \draw[line] (0.4,-0.7) -- (0.4,-1.2);

    \draw[thickline] (0.5,0.0) -- (0.5,-0.5);
    \draw[thickline] (0.5,-0.7) -- (0.5,-1.2);
    
    \node[M] at (0,0) {\small $M_0$};

    \node[M] at (0,-1.2) {\small $\tilde M_0$};

    \node[] at (1.2,-0.6) {\small $\approx$};

  \end{scope}
  
  \begin{scope}[shift={(2.4,0.4)}]

    \foreach \i in {-0.5,0.5} {
        \draw[thickline] (\i,0.0) -- (\i,-0.5);
        \draw[thickline] (\i,-1.5) -- (\i,-2.0);
    }

    \foreach \i in {-0.4,0.4} {
        \draw[line] (\i,0.0) -- (\i,-0.5);
        \draw[line] (\i,-1.5) -- (\i,-2.0);
    }
    
    \node[M] at (0,0) {\small $M_0$};

    \node[M] at (0,-2) {\small $\tilde M_0$};

    \draw[thickline] (-0.45,-0.65) -- (-0.45,-1.35);

    \node[proj, shape border rotate=180] at (-0.45,-0.65) {\small $P$};

    \node[proj] at (-0.45,-1.35) {\small $Q$};

  \end{scope}

\end{tikzpicture} \, , \nonumber
\end{equation}
which is equivalent to optimizing the Frobenius norm
\begin{equation}
\min_{\mathrm{rank}(PQ) \leq \chi} \Big| {M}_0^\dagger (I - PQ) \tilde{{M}}_0 \Big|^2.
\label{gs_FrobeniusNorm}
\end{equation}

By the Eckart-Young-Mirsky theorem~\cite{Eckart_Young_1936,Mirsky1960}, this optimization problem can be cast as performing an SVD on $M_0^\dagger \tilde M_0$ and retaining the largest $\chi$ singular values, 
\begin{equation}
    M_0^\dagger \tilde M_0 \approx U  \Lambda V^\dagger,
\end{equation}
to define the truncated $\Lambda$. The optimal projectors are given from the singular vectors by
\begin{equation}
P = \tilde M_0 {V} {\Lambda}^{-1/2}, \quad Q = {\Lambda}^{-1/2} {U}^\dagger M_0^\dagger.
\label{projector}
\end{equation}
The RG processes are applied cyclically in all four directions until convergence is reached.

Once the CTM tensors have converged, the expectation values of any local observable (ground-state energy, magnetization, \dots) can be calculated by contracting a simple, finite tensor network as represented in Eq.~\eqref{ctm_env}. The accuracy of the expectation values obtained by contracting the infinite tensor network using CTMRG is controlled by the rank of the projectors $P$ and $Q$, i.e.~by the CTM bond dimension, $\chi$.

\subsection{Excitation Calculation by GS-CTMRG}

To compute the excitation spectrum of a quantum many-body system, we require the zero-temperature dynamical structure factor
\begin{equation}
\begin{split}
&S^{\alpha \beta}(\mathbf{k},\omega) = \bra{\Psi_0} S^{\alpha}_{-k} \delta(\omega - H + E_0) S^{\beta}_{k}\ket{\Psi_0} \\
=& \sum_{m} \bra{\Psi_0} S^{\alpha}_{-k} \ket{m} \bra{m}S^{\beta}_{k}\ket{\Psi_0} \delta(\omega - E_m + E_0).
\end{split}
\label{DSF}
\end{equation}
Within the tensor-network representation, once an accurate ground-state iPEPS, $\ket{\Psi_0}$, has been obtained, the excited state $\ket{m}$, a single quasiparticle localized in momentum space, can be described approximately by a perturbation of $\ket{\Psi_0}$~\cite{Vanderstraeten2015}. This excited state is constructed by replacing one local ground-state tensor ($A$) of $\ket{\Psi_0}$ at site {\bf r} by an excitation tensor $B$, 
\begin{equation}
  \begin{tikzpicture}[scale=0.8, transform shape]
    
    \begin{scope}
        \node[anchor=east, font=\large] at (-1.5, -1) {$\ket{\Phi_r(B)} =$};
        
        \foreach \i in {0,1,2} {
            \draw[line] (\i, 0.5) -- (\i, -2.5);
            \draw[line] (-0.5, -\i) -- (2.5, -\i);
        }

        \foreach \x in {0,1,2} {
            \foreach \y in {0,1,2} {
                \draw[line] (\x, -\y) -- (\x+0.4, -\y-0.3);
                \node[A] at (\x, -\y) {$A$};
            }
        }

        \node[A, B] at (1, -1) {$B$};

        \node[] at (1.4, -0.7) {$r$};

        \node at (3, -1) {$\dots$};
        \node at (-1, -1) {$\dots$};
        \node at (1, 1.0) {$\vdots$};
        \node at (1, -3.0) {$\vdots$};
    \end{scope}

\end{tikzpicture}. \nonumber
\end{equation}
which can be interpreted as creating a quasiparticle (wave packet) that is localized in space. To restore translational invariance, all such local wave packets are superposed with appropriate phase factors, $\exp(i\mathbf{k}\cdot\mathbf{r})$, to yield a Bloch state with well-defined momentum {\bf k},
\begin{equation}
\ket{m} = \ket{\Phi_{k}(B_m)} =\sum_{r} e^{i\mathbf{k}\cdot \mathbf r}\ket{\Phi_{r} (B_m)}.
\end{equation}

For a given {\bf k}, the corresponding excitation Ansatz $\ket{m}$ is parametrized by the tensor $B_m$, which is determined variationally. In practice, one first imposes the orthogonality condition
\begin{equation}
\bra{\Phi_{k} (B_m)} \Psi_0 \rangle = 0 
\end{equation}
and thereby constructs a set of basis tensors $\{B_m\}$ spanning the tangent space of the ground state. Because tensor-network states are highly redundant, these tangent-space basis states are not linearly independent. To quantify this statement, we note first that $\ket{\Phi_{k}(B)}$ vanishes identically when $B$ takes the forms \begin{equation}
  \begin{tikzpicture}[scale=0.8, transform shape]

  \begin{scope}
  \begin{scope}
    \draw[line] (0,0) -- (0,-0.5);
    
    \draw[line] (-0.5,0) -- (0.5,0);
    \draw[line] (-0.4,-0.3) -- (0.4,0.3);
    
    \node[A, B] at (0,0) {\small $B_X$};

    \node[] at (1,0) {$=$};
    \node[] at (2.0,0) {$e^{i k_x}$};
    
  \end{scope}

  \begin{scope}[shift={(3.0,0)}]
    \draw[line] (0,0) -- (0,-0.5);
    
    \draw[line] (-0.5,0) -- (1.5,0);
    \draw[line] (-0.4,-0.3) -- (0.4,0.3);
    
    \node[A] at (0,0) {\small $A$};
    \node[A, rectangle] at (1,0) {\small $X$};

    \node[] at (2,0) {$-$};
    
  \end{scope}

  \begin{scope}[shift={(6.0,0)}]
    \draw[line] (1,0) -- (1,-0.5);
    
    \draw[line] (-0.5,0) -- (1.5,0);
    \draw[line] (0.6,-0.3) -- (1.4,0.3);
    
    \node[A, rectangle] at (0,0) {\small $X$};
    \node[A] at (1,0) {\small $A$};
    
  \end{scope}
  \end{scope}

  \begin{scope}[shift={(0, -2.5)}]
  \begin{scope}
    \draw[line] (0,0) -- (0,-0.5);
    
    \draw[line] (-0.5,0) -- (0.5,0);
    \draw[line] (-0.4,-0.3) -- (0.4,0.3);
    
    \node[A, B] at (0,0) {\small $B_Y$};

    \node[] at (1,0) {$=$};
    \node[] at (2.0,0) {$e^{i k_y}$};
    
  \end{scope}

  \begin{scope}[shift={(3.5,0)}]
    \draw[line] (0,0) -- (0,-0.5);
    
    \draw[line] (-0.5,0) -- (0.5,0);
    \draw[line] (-0.4*3.2,-0.3*3.2) -- (0.4,0.3);
    
    \node[A] at (0,0) {\small $A$};
    \node[A, rectangle] at (-0.4*2,-0.3*2) {\small $Y$};

    \node[] at (1.5,0) {$-$};
    
  \end{scope}

  \begin{scope}[shift={(5,0)}]
    \draw[line] (1,0) -- (1,-0.5);
    
    \draw[line] (0.5,0) -- (1.5,0);
    \draw[line] (0.6,-0.3) -- (1+0.4*3.2,0.3*3.2);
    
    \node[A, rectangle] at (1+0.4*2,0.3*2) {\small $Y$};
    \node[A] at (1,0) {\small $A$};
    
  \end{scope}
  \end{scope}

\end{tikzpicture}, \nonumber
\end{equation}
where $X$ and $Y$ are arbitrary $D$$\times$$D$ matrices. This means that there exist $(2D^2 - 1)$ null modes that leave the iPEPS wave function invariant, where the $-1$ arises because $X = I$ and $Y = I$ correspond to the ground state, and hence that there are $dD^4 - (2D^2 - 1)$ linearly independent excitation basis states.

Further, the remaining independent basis states, $\{\ket{\Phi_{k}(B_m)} \}$, are not mutually orthogonal, which mandates the computation not only of the effective Hamiltonian matrix,
\begin{equation}
    \mathcal{H}^\text{eff}_{mn} = \langle \Phi_{k} (B^\dagger_m)| (H - E_0) |\Phi_{k} (B_n) \rangle,
    \label{eq:hmn}
\end{equation} 
but also of the norm matrix,
\begin{equation}
    \mathcal{N}^\text{eff}_{mn} = \langle \Phi_{k} (B^\dagger_m)| \Phi_{k} (B_n) \rangle,
    \label{eq:Nmn}
\end{equation}
in this basis. One then solves the generalized eigenvalue problem
\begin{equation}
\sum_n \mathcal{H}^\text{eff}_{mn} v_{np} = \sum_n E_p \mathcal{N}^\text{eff}_{mn} v_{np},
\label{GEVP}
\end{equation}
to determine the final excitation tensor, 
\begin{equation}
    \tilde B_p = \sum_n B_n v_{np},
\end{equation}
and the corresponding excitation energy, $E_p$. With these results one may finally evaluate the dynamical structure factor of Eq.~\eqref{DSF}.

The central technical challenge in iPEPS spectral calculations is therefore the accurate and efficient evaluation of the matrix elements of the effective Hamiltonian \eqref{eq:hmn},
\begin{equation}
    \mathcal{H}^\text{eff}_{mn} = \sum_{r, r', r''} e^{i\mathbf{k} \cdot (\mathbf{r} - \mathbf{r}')} \langle \Phi_{r'}(B^\dagger_m)| h_{r''} |\Phi_{r}(B_n) \rangle,
\end{equation} 
where $h_{r''}$ is the local Hamiltonian at site {\bf r}$''$ minus the ground-state energy per site. This triple infinite lattice summation can be simplified by exploiting translational invariance, fixing the local Hamiltonian at the center to eliminate one of the sums,
\begin{equation}
    \mathcal{H}^\text{eff}_{mn} = \sum_{r, r'} e^{i\mathbf{k} \cdot (\mathbf{r} - \mathbf{r}')} \langle \Phi_{r'} (B^\dagger_m)| h_{0} |\Phi_{r}(B_n) \rangle,
\end{equation}
but this double summation remains extremely demanding to compute.

Ponsioen \emph{et al.}~\cite{ponsioen2020,ponsioen2022} overcame this difficulty by combining CTMRG~\cite{CTMRG_1,CTMRG_2,CTMRG_3} with AD~\cite{Liao2019}, rewriting the double infinite sum in terms of a CTM summation, where the contributions of all possible positions of the excitation tensors are accumulated during the CTM renormalization
process. To illustrate this by example, we introduce a new CTM tensor, $BE_4$, to represent all contributions where the excitation tensor ($B$) is located to the left of the central site,
\begin{equation}
  \begin{tikzpicture}[scale=0.8, transform shape]

  \begin{scope}
    \foreach \i in {0,2} {
        \draw[thickline] (\i,0) -- (\i,-2);
        \draw[thickline] (0,-\i) -- (2,-\i);
    }

    \draw[line] (1,0) -- (1,-2);
    \draw[line] (0,-1) -- (2,-1);
    
    \node[E] at (0,0) {\small $C_1$};
    \node[E] at (2,0) {\small $C_2$};
    \node[E] at (2,-2) {\small $C_3$};
    \node[E] at (0,-2) {\small $C_4$};
    \node[E] at (1,0) {\small $E_1$};
    \node[E] at (2,-1) {\small $E_2$};
    \node[E] at (1,-2) {\small $E_3$};
    \node[E, B] at (0,-1) {\small $BE_4$};
    
    \node[A, draw=white] at (1,-1) {}; 
    
  \end{scope}

  \begin{scope}[shift={(5.5,0)}]
    
    \begin{scope}
      \foreach \i in {0,...,3} \draw[line] (\i,0.5) -- (\i,-2.5);
      \foreach \i in {0,1,2} \draw[line] (-0.5,-\i) -- (3.5,-\i);
      
      \foreach \x in {0,...,3}
        \foreach \y in {0,1,2} {
          \pgfmathtruncatemacro{\skipnode}{(\x == 1 && \y == 1) || (\x == 2 && \y == 1) ? 1 : 0}
          \ifnum\skipnode=0
            \node[A] at (\x,-\y) {\small $a$};
          \fi
        }
      
      \node[A, B] at (1,-1) {\small $b$};
      \node[A, draw=white] at (2,-1) {}; 
      
      \node at (-2.5,-1) {$\approx$};
      \node at (-1.5,-1) {$e^{-ik_x}$};
      \node at (-0.75,-1) {$\dots$};
      \node at (3.8,-1) {$\dots$};
      \node at (2,0.8) {$\vdots$};
      \node at (2,-2.8) {$\vdots$};
    \end{scope}

    \begin{scope}[shift={(0,-4.2)}]
      \foreach \i in {0,...,3} \draw[line] (\i,0.5) -- (\i,-2.5);
      \foreach \i in {0,1,2} \draw[line] (-0.5,-\i) -- (3.5,-\i);
      
      \foreach \x in {0,...,3}
        \foreach \y in {0,1,2} {
          \pgfmathtruncatemacro{\skipnode}{(\x == 0 && \y == 1) || (\x == 2 && \y == 1) ? 1 : 0}
          \ifnum\skipnode=0
            \node[A] at (\x,-\y) {\small $a$};
          \fi
        }
      
      \node[A, B] at (0,-1) {\small $b$};
      \node[A, draw=white] at (2,-1) {};

      \node at (-2.5,-1) {$+$};
      \node at (-1.5,-1) {$e^{-2ik_x}$};
      \node at (-0.75,-1) {$\dots$};
      \node at (3.8,-1) {$\dots$};
      \node at (2,0.8) {$\vdots$};
      \node at (2,-2.8) {$\vdots$};
    \end{scope}

    \begin{scope}[shift={(0,-6.5)}]
      \node at (-2.5,-1) {$+$};
      \node at (-1.8,-1) {$\dots$};
    \end{scope}
    
  \end{scope}

\end{tikzpicture}. \nonumber
\end{equation}
Here we have defined new double-layer tensors, $b$, $b^{\dagger}$, and $bb^{\dagger}$, containing the single-site $B$ tensors, using different colors to distinguish between tensors containing different types of excitation,
\begin{equation}
  \begin{tikzpicture}[scale=0.8, transform shape]
  \begin{scope}
    \draw[line] (-0.5,0) -- (0.5,0);
    \draw[line] (0,-0.5) -- (0,0.5);
    \node[A, B] at (0,0) {\small $b$};
    
    \node at (1.0,0) {\large $=$};
  \end{scope}

  \begin{scope}[shift={(2.0, 0.5)}]
    \draw[line] (0,0) -- (0,-1);
    
    \foreach \y in {0, -1} {
        \draw[line] (-0.5,\y) -- (0.5,\y);
        \draw[line] (-0.5,\y-0.3) -- (0.5,\y+0.3);
    }
    
    \node[A, B] at (0,0) {\small $B$};
    \node[A] at (0,-1) {\small $A^\dagger$};
  \end{scope}

  \begin{scope}[shift={(3.8, 0.0)}]
    \draw[line] (-0.5,0) -- (0.5,0);
    \draw[line] (0,-0.5) -- (0,0.5);
    \node[A, Bd] at (0,0) {\small $b^\dagger$};
    
    \node at (1.0,0) {\large $=$};
  \end{scope}

  \begin{scope}[shift={(5.8, 0.5)}]
    \draw[line] (0,0) -- (0,-1);
    
    \foreach \y in {0, -1} {
        \draw[line] (-0.5,\y) -- (0.5,\y);
        \draw[line] (-0.5,\y-0.3) -- (0.5,\y+0.3);
    }
    
    \node[A] at (0,0) {\small $A$};
    \node[A, Bd] at (0,-1) {\small $B^\dagger$};
  \end{scope}

  \begin{scope}[shift={(7.6, 0.0)}]
    \draw[line] (-0.5,0) -- (0.5,0);
    \draw[line] (0,-0.5) -- (0,0.5);
    \node[A, BBd] at (0,0) {\small $bb^\dagger$};
    
    \node at (1.0,0) {\large $=$};
  \end{scope}

  \begin{scope}[shift={(9.6, 0.5)}]
    \draw[line] (0,0) -- (0,-1);
    
    \foreach \y in {0, -1} {
        \draw[line] (-0.5,\y) -- (0.5,\y);
        \draw[line] (-0.5,\y-0.3) -- (0.5,\y+0.3);
    }
    
    \node[A, B] at (0,0) {\small $B$};
    \node[A, Bd] at (0,-1) {\small $B^\dagger$};
  \end{scope}

\end{tikzpicture}. \nonumber
\end{equation}
In the same manner, we introduce the boundary tensors $\{BC_i,BE_i\}$ that contain $B$, $\{B^\dagger C_i,B^\dagger E_i\}$ that contain $B^\dagger$, and $\{BB^\dagger C_i, BB^\dagger E_i\}$ that contain both $B$ and $B^\dagger$.

To compute these new boundary tensors, one sums over all possible positions of the excitation tensor concurrently in each CTMRG iteration. Using again the edge tensor $BE_4$, its update involves the two terms
\begin{equation}
  \begin{tikzpicture}[scale=0.8, transform shape]

  \begin{scope}
    \draw[line] (0,0) -- (0.7,0);
    \draw[thickline] (0,-0.7) -- (0,0.7);
    \node[E, B] at (0,0) {\small $BE_4'$};
    \node at (1.4,0) {$=~\Bigg($};
  \end{scope}

  \begin{scope}[shift={(2.2,0)}]
    \draw[thickline] (0,-0.4) -- (0,0.4);
    \draw[line] (0,0) -- (1.7,0);
    \draw[line] (1,-0.4) -- (1,0.4);
    \node[E, B] at (0,0) {\small $BE_4$};
    \node[A] at (1,0) {\small $a$};
    \node at (2.1,0) {$+$};
    \pic at (0,0) {P};
    \pic at (0,0) {Q};
  \end{scope}

  \begin{scope}[shift={(5.1,0)}]
    \draw[thickline] (0,-0.4) -- (0,0.4);
    \draw[line] (0,0) -- (1.7,0);
    \draw[line] (1,-0.4) -- (1,0.4);
    \node[E] at (0,0) {\small $E_4$};
    \node[A, B] at (1,0) {\small $b$};
    \node at (1.75,0) {$~\Bigg)$};
    \node at (2.5,0) {$\cdot~e^{-i k_x}$};
    \pic at (0,0) {P};
    \pic at (0,0) {Q};
  \end{scope}

\end{tikzpicture}, \nonumber
\end{equation}
where a phase factor is included to recover the momentum-dependent phase of the original expression. Similarly, for $B^\dagger E_4$
\begin{equation}
  \begin{tikzpicture}[scale=0.8, transform shape]

  \begin{scope}
    \draw[line] (0,0) -- (0.7,0);
    \draw[thickline] (0,-0.7) -- (0,0.7);
    \node[E, Bd] at (0,0) {\small $B^\dagger E_4'$};
    \node at (1.4,0) {$=~\Bigg($};
  \end{scope}

  \begin{scope}[shift={(2.2,0)}]
    \draw[thickline] (0,-0.4) -- (0,0.4);
    \draw[line] (0,0) -- (1.7,0);
    \draw[line] (1,-0.4) -- (1,0.4);
    \node[E, Bd] at (0,0) {\small $B^\dagger E_4$};
    \node[A] at (1,0) {\small $a$};
    \node at (2.1,0) {$+$};
    \pic at (0,0) {P};
    \pic at (0,0) {Q};
  \end{scope}

  \begin{scope}[shift={(5.1,0)}]
    \draw[thickline] (0,-0.4) -- (0,0.4);
    \draw[line] (0,0) -- (1.7,0);
    \draw[line] (1,-0.4) -- (1,0.4);
    \node[E] at (0,0) {\small $E_4$};
    \node[A, Bd] at (1,0) {\small $b^\dagger $};
    \node at (1.75,0) {$~\Bigg)$};
    \node at (2.5,0) {$\cdot~e^{+i k_x}$};
    \pic at (0,0) {P};
    \pic at (0,0) {Q};
  \end{scope}

\end{tikzpicture}, \nonumber
\end{equation}
while the update of $BB^\dagger E_4$ involves the four terms
\begin{equation}
  \begin{tikzpicture}[scale=0.8, transform shape]

  \begin{scope}
    \draw[line] (0,0) -- (0.9,0);
    \draw[thickline] (0,-0.7) -- (0,0.7);
    \node[E, BBd] at (0,0) {\small $BB^\dagger E_4'$};
    \node at (1.25,0) {$=$};
  \end{scope}

  \begin{scope}[shift={(2.25,0)}]
    \draw[thickline] (0,-0.4) -- (0,0.4);
    \draw[line] (0,0) -- (1.7,0);
    \draw[line] (1,-0.4) -- (1,0.4);
    \node[E, BBd] at (0,0) {\small $BB^\dagger E_4$};
    \node[A] at (1,0) {\small $a$};
    \node at (2.1,0) {$+$};
    \pic at (0,0) {P};
    \pic at (0,0) {Q};
  \end{scope}

  \begin{scope}[shift={(5.25,0)}]
    \draw[thickline] (0,-0.4) -- (0,0.4);
    \draw[line] (0,0) -- (1.7,0);
    \draw[line] (1,-0.4) -- (1,0.4);
    \node[E, B] at (0,0) {\small $BE_4$};
    \node[A, Bd] at (1,0) {\small $b^\dagger$};
    \pic at (0,0) {P};
    \pic at (0,0) {Q};
  \end{scope}

  \begin{scope}[shift={(2.25,-3)}]
    \node at (-1,0) {$+$};
    \draw[thickline] (0,-0.4) -- (0,0.4);
    \draw[line] (0,0) -- (1.7,0);
    \draw[line] (1,-0.4) -- (1,0.4);
    \node[E, Bd] at (0,0) {\small $B^\dagger E_4$};
    \node[A, B] at (1,0) {\small $b$};
    \node at (2.1,0) {$+$};
    \pic at (0,0) {P};
    \pic at (0,0) {Q};
  \end{scope}

  \begin{scope}[shift={(5.25,-3)}]
    \draw[thickline] (0,-0.4) -- (0,0.4);
    \draw[line] (0,0) -- (1.7,0);
    \draw[line] (1,-0.4) -- (1,0.4);
    \node[E] at (0,0) {\small $E_4$};
    \node[A, BBd] at (1,0) {\small $bb^\dagger$};
    \pic at (0,0) {P};
    \pic at (0,0) {Q};
  \end{scope}

\end{tikzpicture}, \nonumber
\end{equation}
where the phase factors cancel. The update rules for the corner tensors are similar, which we illustrate with the left move for $BC_1$, given by
\begin{equation}
  \begin{tikzpicture}[scale=0.8, transform shape]

  \begin{scope}
    \draw[thickline] (0,0) -- (0,-0.7);
    \draw[thickline] (0,0) -- (0.7,0);
    \node[E, B] at (0,0) {\small $BC_1'$};
    \node at (1.4,0) {$=~\Bigg($};
  \end{scope}

  \begin{scope}[shift={(2.1,0)}]
    \draw[thickline] (0,0) -- (0,-0.4);
    \draw[thickline] (0,0) -- (1.7,0);
    \draw[line] (1,0) -- (1,-0.4);
    \node[E, B] at (0,0) {\small $BC_1$};
    \node[E] at (1,0) {\small $E_1$};
    \pic at (0,0) {P};
    \node at (2.1,0) {$+$};
  \end{scope}

  \begin{scope}[shift={(5,0)}]
    \draw[thickline] (0,0) -- (0,-0.4);
    \draw[thickline] (0,0) -- (1.7,0);
    \draw[line] (1,0) -- (1,-0.4);
    \node[E] at (0,0) {\small $C_1$};
    \node[E, B] at (1,0) {\small $BE_1$};
    \node at (1.75,0) {$~\Bigg)$};
    \node at (2.5,0) {$\cdot~e^{-i k_x}$};
    \pic at (0,0) {P};
  \end{scope}

\end{tikzpicture}.
\end{equation}

Once all of the CTM tensors have been obtained, the effective Hamiltonian \eqref{eq:hmn} and norm matrices \eqref{eq:Nmn} can be evaluated. Taking the example of a general local Hamiltonian $h$ supported on a 2$\times$2 plaquette, its expectation value is computed by the summation
\begin{equation}
  \begin{tikzpicture}[scale=0.8, transform shape]

  \begin{scope}[shift={(0,1.5)}]
    \node at (0,0) {$\sum_{r_1,r_2} e^{i k \cdot (-r_1 + r_2)} \mel{\Phi_{r_1}(B^{\dagger})}{h}{\Phi_{r_2}(B)}=$};
  \end{scope}

  \begin{scope}[shift={(-3,0)}]
    \node[H] at (1.5, -1.5) {\small $h$};
    \foreach \i in {0, 3} {
        \draw[thickline] (\i,0) -- (\i,-3);
        \draw[thickline] (0,-\i) -- (3,-\i);
    }

    \foreach \i in {1, 2} {
        \draw[line] (\i,0) -- (\i,-3);
        \draw[line] (0,-\i) -- (3,-\i);
    }
    
    \node[E, BBd] at (0,0) {\small $BB^\dagger C_1$};
    \node[E] at (3,0) {\small $C_2$};
    \node[E] at (3,-3) {\small $C_3$};
    \node[E] at (0,-3) {\small $C_4$};
    \node[E] at (1,0) {\small $E_1$};
    \node[E] at (2,0) {\small $E_1$};
    \node[E] at (3,-1) {\small $E_2$};
    \node[E] at (3,-2) {\small $E_2$};
    \node[E] at (1,-3) {\small $E_3$};
    \node[E] at (2,-3) {\small $E_3$};
    \node[E] at (0,-2) {\small $E_4$};
    \node[E] at (0,-1) {\small $E_4$};
    
    \foreach \x in {1,2}
        \foreach \y in {1,2}
            \node[A] at (\x,-\y) {\small $a$};
            
    \node at (4,-1.5) {$+$};
  \end{scope}

  \begin{scope}[shift={(2,0)}]
    \node[H] at (1.5, -1.5) {\small $h$};
    \foreach \i in {0, 3} {
        \draw[thickline] (\i,0) -- (\i,-3);
        \draw[thickline] (0,-\i) -- (3,-\i);
    }

    \foreach \i in {1, 2} {
        \draw[line] (\i,0) -- (\i,-3);
        \draw[line] (0,-\i) -- (3,-\i);
    }
    
    \node[E, B] at (0,0) {\small $BC_1$};
    \node[E] at (3,0) {\small $C_2$};
    \node[E] at (3,-3) {\small $C_3$};
    \node[E] at (0,-3) {\small $C_4$};
    \node[E] at (1,0) {\small $E_1$};
    \node[E] at (2,0) {\small $E_1$};
    \node[E] at (3,-1) {\small $E_2$};
    \node[E] at (3,-2) {\small $E_2$};
    \node[E] at (1,-3) {\small $E_3$};
    \node[E] at (2,-3) {\small $E_3$};
    \node[E] at (0,-2) {\small $E_4$};
    \node[E, Bd] at (0,-1) {\small $B^\dagger E_4$};
    
    \foreach \x in {1,2}
        \foreach \y in {1,2}
            \node[A] at (\x,-\y) {\small $a$};
            
    \node at (4,-1.5) {$+$};
    \node at (5,-1.5) {$\cdots$};
  \end{scope}

\end{tikzpicture}. \nonumber
\end{equation}
To execute these summation steps, Ponsioen \emph{et al.}~\cite{ponsioen2020,ponsioen2022} employed the ground-state projectors $P$ and $Q$ of Eq.~(\ref{projector}) to update the CTM tensors. While an eminently reasonable proposal using well defined projector matrices, this turns out to be the primary source of error in constructing accurate effective Hamiltonian and norm matrices for the stable and accurate computation of excitation spectra. We therefore turn to the excitation-tailored projector construction method with which we replace this approach. 

\section{Excitation Calculation by ET-CTMRG}
\label{secs2}

\subsection{Projector Construction for ET-CTMRG}

Based on the introduction of Sec.~\ref{secs1}, the modifications constituting the ET-CTMRG method can be stated rather succinctly. To recap, in the GS-CTMRG method~\cite{ponsioen2020,ponsioen2022}, the CTM boundary tensors that contain excitation tensors ($B$, $B^\dagger$, both $B$ and $B^\dagger$) are truncated using projectors $P$ and $Q$ built only from the ground-state half-blocks $M_0$ and $\tilde{M}_0$. These projectors are optimized to minimize the truncation error for the ground-state configuration alone, not to capture the subspace relevant to configurations with excitation tensors. As a result, the effective Hamiltonian and norm matrices computed in the CTM summation acquire systematic errors, which can lead to numerical instabilities when solving the generalized eigenvalue problem of Eq.~(\ref{GEVP}), including unphysical negative excitation energies.

To remedy the basic shortcoming, we introduce a new projector construction that incorporates information from all relevant half-block configurations both with and without excitation tensors. Following the GS-CTMRG scheme, we expand the definition of the cost function for constructing projectors, Eq.~(\ref{gs_FrobeniusNorm}), to include all relevant half-block configurations,
\begin{equation}
\min_{\mathrm{rank}(X) \leq \chi} \sum_{i,j=0}^{3} \Big| M_i^\dagger (I - X) \tilde{M}_j \Big|^2,
\label{exci-FrobeniusNorm}
\end{equation}
where $X = PQ$ and the new tensors $M_1$ and $\tilde{M}_1$ denote half-blocks with a single $B$ excitation tensor [Fig.~1(b) of the main text], $M_2$ and $\tilde{M}_2$ denote half-blocks with a single $B^\dagger$, and $M_3$ and $\tilde{M}_3$ denote half-blocks containing both $B$ and $B^\dagger$. As an example we illustrate the construction of $M_1$, 
\begin{equation}
    \begin{tikzpicture}[scale=0.7, transform shape]

  \begin{scope}

    \foreach \i in {-2.5,-1.5} {
        \draw[thickline] (\i,-0.5) -- (\i,-1.0);
    }

    \foreach \i in {-2.4,-1.6} {
        \draw[line] (\i,-0.5) -- (\i,-1.0);
    }
    
    \node[M, B] at (-2.0,-0.5) {\small $M_1$};

  \end{scope}

  \begin{scope}

    \node[] at (-0.9,-0.5) {$\equiv$};
    
    \foreach \i in {0,3} {
        \draw[thickline] (\i,0) -- (\i,-1.6);
    }

    \foreach \i in {1,2} {
        \draw[line] (\i,0) -- (\i,-1.6);
    }

    \draw[thickline] (0,0) -- (3,0);
    \draw[line] (0,-1) -- (3,-1);
    
    \node[E,B] at (0,0) {\small $BC_1$};
    \node[E] at (3,0) {\small $C_2$};
    \node[E] at (1,0) {\small $E_1$};
    \node[E] at (2,0) {\small $E_1$};
    \node[E] at (3,-1) {\small $E_2$};
    \node[E] at (0,-1) {\small $E_4$};
    \node[A] at (1,-1) {\small $a$};     
    \node[A] at (2,-1) {\small $a$};

    \node[] at (3.8,-0.5) {$+$};

  \end{scope}

  \begin{scope}[shift={(4.6,0)}]
    
    \foreach \i in {0,3} {
        \draw[thickline] (\i,0) -- (\i,-1.6);
    }

    \foreach \i in {1,2} {
        \draw[line] (\i,0) -- (\i,-1.6);
    }

    \draw[thickline] (0,0) -- (3,0);
    \draw[line] (0,-1) -- (3,-1);
    
    \node[E] at (0,0) {\small $C_1$};
    \node[E] at (3,0) {\small $C_2$};
    \node[E] at (1,0) {\small $E_1$};
    \node[E] at (2,0) {\small $E_1$};
    \node[E] at (3,-1) {\small $E_2$};
    \node[E,B] at (0,-1) {\small $BE_4$};
    \node[A] at (1,-1) {\small $a$};
    \node[A] at (2,-1) {\small $a$};

  \end{scope}

  \begin{scope}[shift={(0.0,-2.5)}]

    \node[] at (-0.9,-0.5) {$+$};
    
    \foreach \i in {0,3} {
        \draw[thickline] (\i,0) -- (\i,-1.6);
    }

    \foreach \i in {1,2} {
        \draw[line] (\i,0) -- (\i,-1.6);
    }

    \draw[thickline] (0,0) -- (3,0);
    \draw[line] (0,-1) -- (3,-1);
    
    \node[E] at (0,0) {\small $C_1$};
    \node[E] at (3,0) {\small $C_2$};
    \node[E,B] at (1,0) {\small $BE_1$};
    \node[E] at (2,0) {\small $E_1$};
    \node[E] at (3,-1) {\small $E_2$};
    \node[E] at (0,-1) {\small $E_4$};
    \node[A] at (1,-1) {\small $a$};
    \node[A] at (2,-1) {\small $a$};

    \node[] at (3.8,-0.5) {$+$};

  \end{scope}

  \begin{scope}[shift={(4.6,-2.5)}]
    
    \foreach \i in {0,3} {
        \draw[thickline] (\i,0) -- (\i,-1.6);
    }

    \foreach \i in {1,2} {
        \draw[line] (\i,0) -- (\i,-1.6);
    }

    \draw[thickline] (0,0) -- (3,0);
    \draw[line] (0,-1) -- (3,-1);
    
    \node[E] at (0,0) {\small $C_1$};
    \node[E] at (3,0) {\small $C_2$};
    \node[E] at (1,0) {\small $E_1$};
    \node[E] at (2,0) {\small $E_1$};
    \node[E] at (3,-1) {\small $E_2$};
    \node[E] at (0,-1) {\small $E_4$};
    \node[A,B] at (1,-1) {\small $b$};
    \node[A] at (2,-1) {\small $a$};

  \end{scope}

  \begin{scope}[shift={(0.0,-5.0)}]

    \node[] at (-0.9,-0.5) {$+$};
    \node[] at (-0.6, -0.5) {$\Bigg($};
    
    \foreach \i in {0,3} {
        \draw[thickline] (\i,0) -- (\i,-1.6);
    }

    \foreach \i in {1,2} {
        \draw[line] (\i,0) -- (\i,-1.6);
    }

    \draw[thickline] (0,0) -- (3,0);
    \draw[line] (0,-1) -- (3,-1);
    
    \node[E] at (0,0) {\small $C_1$};
    \node[E] at (3,0) {\small $C_2$};
    \node[E] at (1,0) {\small $E_1$};
    \node[E,B] at (2,0) {\small $BE_1$};
    \node[E] at (3,-1) {\small $E_2$};
    \node[E] at (0,-1) {\small $E_4$};
    \node[A] at (1,-1) {\small $a$};     
    \node[A] at (2,-1) {\small $a$};

    \node[] at (3.8,-0.5) {$+$};

  \end{scope}

  \begin{scope}[shift={(4.6,-5.0)}]
    
    \foreach \i in {0,3} {
        \draw[thickline] (\i,0) -- (\i,-1.6);
    }

    \foreach \i in {1,2} {
        \draw[line] (\i,0) -- (\i,-1.6);
    }

    \draw[thickline] (0,0) -- (3,0);
    \draw[line] (0,-1) -- (3,-1);
    
    \node[E] at (0,0) {\small $C_1$};
    \node[E] at (3,0) {\small $C_2$};
    \node[E] at (1,0) {\small $E_1$};
    \node[E] at (2,0) {\small $E_1$};
    \node[E] at (3,-1) {\small $E_2$};
    \node[E] at (0,-1) {\small $E_4$};
    \node[A] at (1,-1) {\small $a$};     
    \node[A,B] at (2,-1) {\small $b$};

  \end{scope}

  \begin{scope}[shift={(0.0,-7.5)}]

    \node[] at (-0.9,-0.5) {$+$};
    
    \foreach \i in {0,3} {
        \draw[thickline] (\i,0) -- (\i,-1.6);
    }

    \foreach \i in {1,2} {
        \draw[line] (\i,0) -- (\i,-1.6);
    }

    \draw[thickline] (0,0) -- (3,0);
    \draw[line] (0,-1) -- (3,-1);
    
    \node[E] at (0,0) {\small $C_1$};
    \node[E,B] at (3,0) {\small $BC_2$};
    \node[E] at (1,0) {\small $E_1$};
    \node[E] at (2,0) {\small $E_1$};
    \node[E] at (3,-1) {\small $E_2$};
    \node[E] at (0,-1) {\small $E_4$};
    \node[A] at (1,-1) {\small $a$};     
    \node[A] at (2,-1) {\small $a$};

    \node[] at (3.8,-0.5) {$+$};

  \end{scope}

  \begin{scope}[shift={(4.6,-7.5)}]
    
    \foreach \i in {0,3} {
        \draw[thickline] (\i,0) -- (\i,-1.6);
    }

    \foreach \i in {1,2} {
        \draw[line] (\i,0) -- (\i,-1.6);
    }

    \draw[thickline] (0,0) -- (3,0);
    \draw[line] (0,-1) -- (3,-1);
    
    \node[E] at (0,0) {\small $C_1$};
    \node[E] at (3,0) {\small $C_2$};
    \node[E] at (1,0) {\small $E_1$};
    \node[E] at (2,0) {\small $E_1$};
    \node[E,B] at (3,-1) {\small $BE_2$};
    \node[E] at (0,-1) {\small $E_4$};
    \node[A] at (1,-1) {\small $a$};     
    \node[A] at (2,-1) {\small $a$};

    \node[] at (3.6, -0.5) {$\Bigg)$};
    \node[] at (4.2, -0.5) {$\cdot~e^{ik_x}$};

  \end{scope}

\end{tikzpicture}, \nonumber
\end{equation}
and all other half-block configurations containing excitation tensors are constructed in a similar manner.

The new cost function [Eq.~(\ref{exci-FrobeniusNorm})] can be further simplified as 
\begin{eqnarray}
\sum_{i,j=0}^{3} && \Big| M_i^\dagger (I - X) \tilde{M}_j \Big|^2 \\
&& = \sum_{i,j=0}^{3} \text{Tr} \, \big[ \tilde{M}_j^\dagger (I - X^\dagger) M_i M_i^\dagger (I - X) \tilde{M}_j \big] \nonumber \\
&& = \sum_{i,j=0}^{3} \text{Tr} \, \big[ \tilde{M}_j \tilde{M}_j^\dagger (I - X^\dagger) M_i M_i^\dagger (I - X) \big] \nonumber \\
&& = \text{Tr} \bigg[ \big( \sum_{j=0}^{3} \tilde{M}_j \tilde{M}_j^\dagger \big) (I - X^\dagger) \big( \sum_{i=0}^{3} M_i M_i^\dagger \big) (I - X) \bigg]. \nonumber
\end{eqnarray}
We next introduce the tensors $K$ and $\tilde{K}$ satisfying the conditions $\sum_{i=0}^{3} M_i M_i^\dagger = K K^\dagger$ and $\sum_{j=0}^{3} \tilde{M}_j \tilde{M}_j^\dagger = \tilde{K} \tilde{K}^\dagger$. This allows the cost function to be reexpressed as 
\begin{eqnarray}
\sum_{i,j=0}^{3} &&\Big| M_i^\dagger (I - X) \tilde{M}_j \Big|^2 \\
&& = \text{Tr} \, [\tilde K \tilde K^\dagger (I - X^\dagger) K K^\dagger (I - X)] \nonumber \\
&& = \text{Tr} \, [\tilde K^\dagger (I - X^\dagger) K K^\dagger (I - X) \tilde K] \nonumber \\
&& = \Big| K^\dagger (I - X) \tilde{K} \Big|^2, \nonumber
\end{eqnarray}
reducing the optimization problem to the same form as in the GS-CTMRG scheme, i.e.~minimizing $\Big|K^\dagger (I - X) \tilde{K}\Big|^2$ under the constraint $\mathrm{rank}(X) \leq \chi$. The optimal $X = PQ$ is obtained from the SVD of $K^\dagger \tilde{K}$, and the projectors $P$ and $Q$ are constructed as in the ground-state case, with $K$ and $\tilde{K}$ replacing $M_0$ and $\tilde{M}_0$, namely
\begin{equation}
  \quad P = \tilde K {V} {\Lambda}^{-1/2}, \quad Q = {\Lambda}^{-1/2} {U}^\dagger K^\dagger ,
\end{equation}
where $U$, $V$, and $\Lambda$ are obtained from the truncated singular-value decomposition of $K^{\dagger} \tilde K$, in which the largest $\chi$ singular values are kept. 

In this way our new projector construction yields a simple and robust formulation. The only extra cost is that of forming the extended half-block matrices, $K$ and $\tilde{K}$, from $\{M_i\}$ and $\{\tilde{M}_j\}$. In practice, the most efficient way to achieve this with minimal numerical errors is by constructing a single matrix $\mathcal{M} = (M_0, \dots, M_3)$ such that $\sum_{i=0}^{3} M_i M_i^\dagger = \mathcal{M} \mathcal{M}^\dagger$. Performing a QR decomposition of $\mathcal{M}^\dagger = QR$ then yields $K = R^\dagger$.

\begin{figure}[tp]
\centering
\includegraphics[width=0.7\columnwidth]{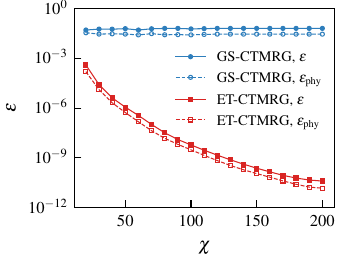}
\caption{Frobenius error, $\varepsilon$, and physical Frobenius error, $\varepsilon_{\text{phy}}$, shown as functions of the CTM bond dimension, $\chi$. Calculations were performed by GS-CTMRG and ET-CTMRG for the square-lattice Heisenberg antiferromagnet (SLHAF) at momentum $(0.2\pi,0.3\pi)$ with $D = 3$.}
\label{fig:square_chi_all_error_convergence}
\end{figure}

\begin{figure}[t]
\centering
\includegraphics[width=0.7\columnwidth]{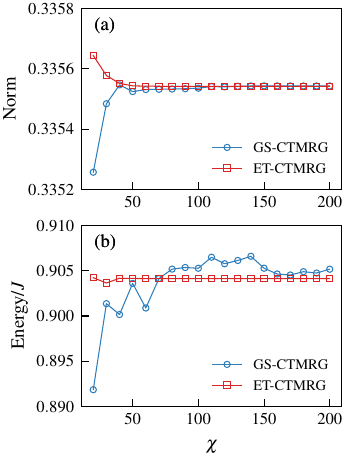}
\caption{(a) Norm and (b) energy of an arbitrarily chosen excited state $B_n$, computed as functions of $\chi$ within GS-CTMRG and ET-CTMRG for the SLHAF at momentum $(0.2\pi,0.3\pi)$ with $D = 3$.}
\label{fig:square_chi_energy_norm_convergence}
\end{figure}

\subsection{Comparing the Convergence of GS-CTMRG and ET-CTMRG with CTM Bond Dimension}

To evaluate the performance of ET-CTMRG against GS-CTMRG, we present a systematic comparison of their convergence behavior with respect to the CTM bond dimension, $\chi$. Consistent with the main text, we define the Frobenius error,
\begin{equation}
 \varepsilon \! = \! \left| K^\dagger (I \! - \! PQ) \tilde K \right| \! = \! \sqrt{\sum_{i,j=0}^{3} \Big| M_i^\dagger (I \! - \! X) \tilde{M}_j \Big|^2}.
\end{equation}
We note that this error contains non-physical components, namely those in which $M$ and $\tilde{M}$ may both contain $B$ simultaneously, or both contain $B^\dagger$. To eliminate these, we define a physical Frobenius error,
\begin{equation}
 \varepsilon_{\text{phy}} = \sqrt{\sum_{(i,j)\in\mathcal{S}} \Big| M_i^\dagger (I - X) \tilde{M}_j \Big|^2},
\end{equation}
where
\begin{equation}
\begin{split}
  \mathcal{S} = \{(0,0),(1,0),(0,1),(2,0),(0,2),\\(3,0),(0,3),(1,2),(2,1)\}  
\end{split}
\end{equation}
 contains all physical combinations of $M_i$ and $\tilde{M}_j$. By definition, $\varepsilon_{\text{phy}}$ must be smaller than $\varepsilon$, and hence minimizing $\varepsilon$ optimizes $\varepsilon_{\text{phy}}$ simultaneously. Figure \ref{fig:square_chi_all_error_convergence} displays the two Frobenius errors as functions of $\chi$ for the square-lattice Heisenberg antiferromagnet (SLHAF), making clear that their behavior is nearly identical: as in Fig.~2 of the main text, within GS-CTMRG there is no clear convergence, whereas within ET-CTMRG the errors converge to extreme precision as $\chi$ is increased. The more compact mathematical form of $\varepsilon$ allows the optimization problem to be cast into a simple SVD, and hence we use this quantity to render ET-CTMRG both practically efficient and straightforward to implement.

 In Fig.~\ref{fig:square_chi_energy_norm_convergence}, we consider the norm [Eq.~\eqref{eq:Nmn}] and energy [Eq.~\eqref{eq:hmn}] of a randomly chosen excited state, $B_n$, at $\mathbf{k} = (0.2\pi,0.3\pi)$. The norm clearly converges significantly faster with increasing $\chi$ under ET-CTMRG than GS-CTMRG [Fig.~\ref{fig:square_chi_energy_norm_convergence}(a)]. The energy converges extremely rapidly with ET-CTMRG, whereas with GS-CTMRG there is no clear sign of convergence [Fig.~\ref{fig:square_chi_energy_norm_convergence}(b)], indicating that GS-CTMRG cannot produce a correctly converged energy simply by increasing $\chi$. The reason is that a $\chi D^2 \to \chi$ truncation is intrinsic to every CTM step, and in GS-CTMRG this inevitably cuts down the subspace relevant to the excited states, regardless of how large $\chi$ is. This also explains why the Frobenius errors fail to converge with $\chi$. The different behavior of the norm and the Hamiltonian matrices stems from the fact that the former involves only a single summation over the excited-state basis, whereas the latter involves double summations, making it much more sensitive to an accurate treatment of the environment.

\begin{figure}[t]
\centering
\includegraphics[width=\columnwidth]{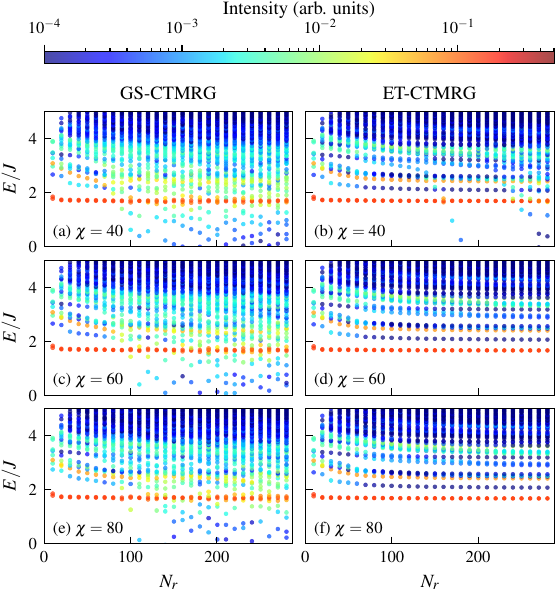}
\caption{Comparison of excitation spectra as functions of the number of retained states, $N_r$. Calculations were performed with GS-CTMRG (a,c,e) and ET-CTMRG (b,d,f) for the SLHAF at momentum $(0.2\pi,0.3\pi)$ with $D = 3$ and $\chi = 40$ (a,b), $60$ (c,d), and $80$ (e,f). The color indicates the spectral weight in each eigenstate.}
\label{fig:square_basis_number_convergence_chi}
\end{figure}

Figure ~\ref{fig:square_basis_number_convergence_chi} studies the stability of the excitation spectrum to the number of retained states [Figs.~3(a,b) of the main text] as $\chi$ is increased. Under GS-CTMRG, increasing $\chi$ yields no pronounced mitigation of the $N_r$-induced energy instability, which is caused by the flawed convergence of the Hamiltonian matrix elements. In contrast, ET-CTMRG achieves considerable stability even at $\chi = 40$, with only minor divergence occurring as $N_r$ is increased; further increasing $\chi$ to $60$ leads to full stability. These results demonstrate an important property of the ET-CTMRG framework that GS-CTMRG fails to deliver, namely that increasing $\chi$ brings a systematic improvement in the stability of the excitation spectrum.

\begin{figure}[t]
\centering
\includegraphics[width=\columnwidth]{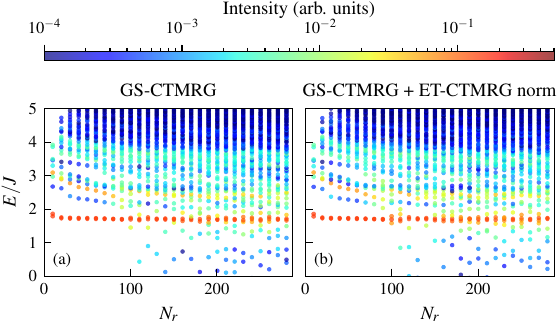}
 \caption{Comparison of $N_r$-induced instabilities in the excitation spectra using GS-CTMRG (a) and using the GS-CTMRG Hamiltonian matrix with the ET-CTMRG norm matrix (b). As in Fig.~\ref{fig:square_basis_number_convergence_chi}, whose panel (c) matches panel (a), calculations were performed for the SLHAF at momentum $(0.2\pi,0.3\pi)$ with $D = 3$ and $\chi = 60$.}
\label{fig:square_basis_number_convergence_vs_norm_exci}
\end{figure}

Historically, the energy instability in GS-CTMRG has often been attributed to an instability of the norm matrix. As further evidence against this conventional wisdom, we build on Figs.~3(c,d) of the main text to compute the spectra for different $N_r$ by substituting the norm matrix calculated by ET-CTMRG into the GS-CTMRG procedure. As Fig.~\ref{fig:square_basis_number_convergence_vs_norm_exci} shows, merely replacing the norm matrix yields no significant mitigation of the instability within GS-CTMRG. This demonstrates clearly that the spectral instability originates primarily from the numerical errors in the Hamiltonian matrix.

\subsection{Properties of ET-CTMRG spectra}

The tensor-network framework is a sophisticated methodology for the efficient representation and truncation of information in a quantum many-body system, and hence iPEPS spectral calculations are performed with a number of variable parameters whose systematic alteration allows internal benchmarking and indicates the convergence of physical observables. The fundamental truncation parameter is the tensor bond dimension, $D$, which cannot be large because the computational cost of spectral calculations scales with $D^{16}$, but even the presently accessible values are sufficient to ensure that the tensor network is representing the correct type of wave function. The CTM bond dimension, $\chi$, is critical for the evaluation of expectation values by CTMRG, and it is usually thought that $\chi > 2D^2$ is sufficient for converged results at each $D$. Here we have shown that larger $\chi$ values within GS-CTMRG do not assist in reducing the truncation error (Figs.~\ref{fig:square_chi_all_error_convergence} and Fig.~2 of the main text), or the norm and energy (Fig.~\ref{fig:square_chi_energy_norm_convergence}), but this is not because the spectrum has converged. The ET-CTMRG procedure makes clear that converging to an accurate spectrum is reflected in the Frobenius error falling monotonically with increasing $\chi$, even to orders of magnitude that have a calculational value, but not necessarily a further physical value given the small $D$. 

As noted in the main text, $N_r$, the number of states retained in representing the excitation spectrum, is an additional parameter in iPEPS spectral calculations. In a system whose spectrum consists of well-defined, discrete $S = 1$ modes arranged sequentially in energy, $N_r$ values of 50-100 can be expected to provide an accurate representation. These values are sufficient to determine that the general form of the spectrum is invariant as $D$ and $\chi$ are altered, and is also invariant across momentum space. In our benchmark example of the SLHAF, all momenta are expected to show well-defined magnon excitations, other than the special point $(\pi,0)$ \cite{DallaPiazza2014,Plumb2014,Powalski2015,Shao2017}, and indeed our calculations corroborate this; for the special point itself, it is unclear that spectra calculated with $D = 4$ are sufficient to discuss the underlying physics, even when the ET-CTMRG method ensures their accurate convergence with $N_r$. 

Nevertheless, any spectrum possessing a branch of ideally noninteracting one-magnon excitations possesses a two-magnon continuum, which is the consequence of scattering processes in which two free magnons absorb the energy and momentum of one scattered probe particle. The nature of the two- and higher-magnon continua in the presence of magnon-magnon interactions, and their effect on the one-magnon branch when it exists, is one of the core questions in quantum magnetism \cite{Stone2005,Verresen2019,hernandez2025}. By its nature, an excitation continuum can only be represented in an iPEPS spectral calculation by retaining very many states, most of which will have a very low spectral weight. An accurate discussion of whether an iPEPS spectrum contains a continuum therefore requires a representation that is stable and convergent to arbitrarily large $N_r$, meaning arbitrarily low spectral weights, and this is a key advantage of ET-CTMRG (Figs.~3 and 4 of the main text, Fig.~\ref{fig:square_basis_number_convergence_chi}, and Figs.~\ref{fig:triangular_basis_convergence} and \ref{fig:kagome_basis_convergence} below). 

One disadvantage is that, although the computational cost of ET-CTMRG is essentially that of GS-CTMRG, it has been shown recently that spectral calculations using GS-CTMRG can be sped up very significantly by storing the projectors $P$ and $Q$. Because these are evaluated using only the ground-state wave function (Sec.~\ref{secs1}B), they need not be reevaluated at every CTMRG step, which reduces the problem from one of repeated SVD operations to a single SVD operation followed by repeated matrix multiplications \cite{kagome_sl}. This speed-up effect increases as $\chi$ is increased, which can place higher $D$ values within calculational reach. The key advantage of ET-CTMRG is that the excited states are included in the tensor environment during the iterative CTMRG process, and this requires the self-consistent inclusion of the spectrum at each step, making the speed-up short-cut explicitly inapplicable. 

\begin{figure}[t]
\centering
\includegraphics[width=0.98\columnwidth]{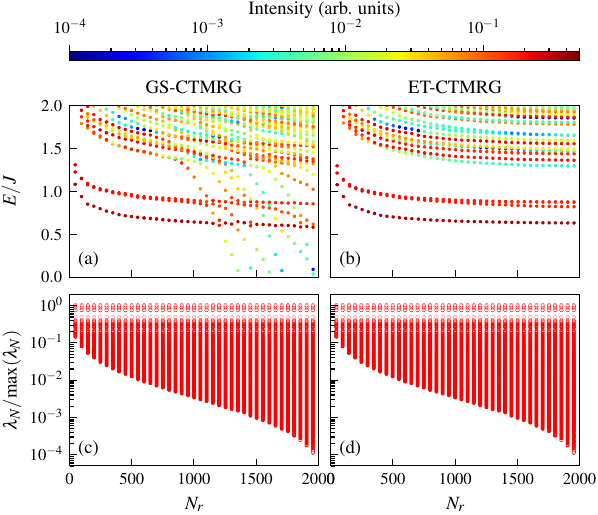}
\caption{Triangular-lattice Heisenberg antiferromagnet: comparison of excitation spectra (a,b) and normalized norm-matrix eigenvalues (c,d) as functions of the number of retained states, $N_r$, obtained with GS-CTMRG (a,c) and ET-CTMRG (b,d), computed at the M point with $D = 4$ and $\chi = 50$. In panels (a) and (b), the color indicates the spectral weight in each eigenstate.}
\label{fig:triangular_basis_convergence}
\end{figure}

\section{Benchmarks for paradigm models}
\label{secs3}

Following the benchmarking of GS-CTMRG and ET-CTMRG shown for the SLHAF in the main text and in Sec.~\ref{secs2}B, here we illustrate their differences for two of the paradigm models in frustrated quantum magnetism, the triangular-lattice Heisenberg antiferromagnet (TLHAF) and the kagome Heisenberg antiferromagnet (KHAF). In the spectrum of the TLHAF at the M point, shown in Fig.~\ref{fig:triangular_basis_convergence}, we observe only a few discrete, low-lying magnonic states below a quasi-continuum that can be ascribed to two- and higher-magnon excitations. We note that the $N_r$ values are considerably larger here than those we used for studying the SLHAF because $D$ is larger. Unlike the SLHAF, the TLHAF spectrum remains rather stable as $N_r$ is increased, although the onset of an instability is clear beyond $N_r = 1000$. At this point only a relatively small number of states becomes unstable, and these continue to change somewhat systematically as $N_r$ is increased. This instability is completely absent in the ET-CTMRG spectrum, which shows systematic convergence to the highest $N_r$. We note that the norm matrix is again nearly identical in GS- and ET-CTMRG, indicating that the origin of the instability in the TLHAF spectrum also lies fully within the effective Hamiltonian matrix. 

\begin{figure}[t]
\centering
\includegraphics[width=0.98\columnwidth]{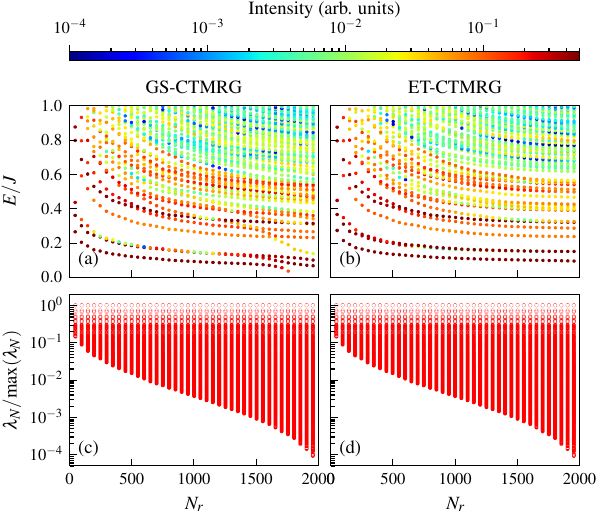}
\caption{Kagome Heisenberg antiferromagnet: comparison of excitation spectra (a,b) and normalized norm-matrix eigenvalues (c,d) as functions of the number of retained states, $N_r$, obtained with GS-CTMRG (a,c) and ET-CTMRG (b,d), computed at the M point with $D = 4$ and $\chi = 50$. In panels (a) and (b), the color indicates the spectral weight in each eigenstate.}
\label{fig:kagome_basis_convergence}
\end{figure}

Turning to the KHAF, in Fig.~\ref{fig:kagome_basis_convergence} we observe a very high density of low-lying states, indicative of an excitation continuum that has descended to the lowest energies (interpreted as the complete deconfinement expected in a U(1) Dirac spin liquid (DSL) \cite{kagome_sl}). In GS-CTMRG, the spectral stability is even higher than for the TLHAF, with only a small number of states becoming unstable towards $N_r = 1500$. As noted in the main text, frustration alone does not seem to be a criterion affecting the stability or otherwise of the GS-CTMRG spectrum. Again the ET-CTMRG calculation has no instability at all, confirming the converged spectrum. Again the norm matrix is not the source of the instability. Figure \ref{fig:kagome_basis_convergence} illustrates why, for the KHAF case, a sped-up GS-CTMRG calculation with $N_r \le 1500$ can be competitive with ET-CTMRG, using its access to higher $D$ with excellent coverage of {\bf k} and the Hamiltonian parameters \cite{kagome_sl} to compensate for the limits in $N_r$. By contrast, this strategy cannot be applied to the TL-XXZ model for K$_2$Co(SeO$_3$)$_2$ because the instabilities are so severe, causing significant differences between the GS- and ET-CTMRG spectra even at $N_r < 50$ (Fig.~4 of the main text). 

\begin{figure*}[t]
\centering
\includegraphics[width=\textwidth]{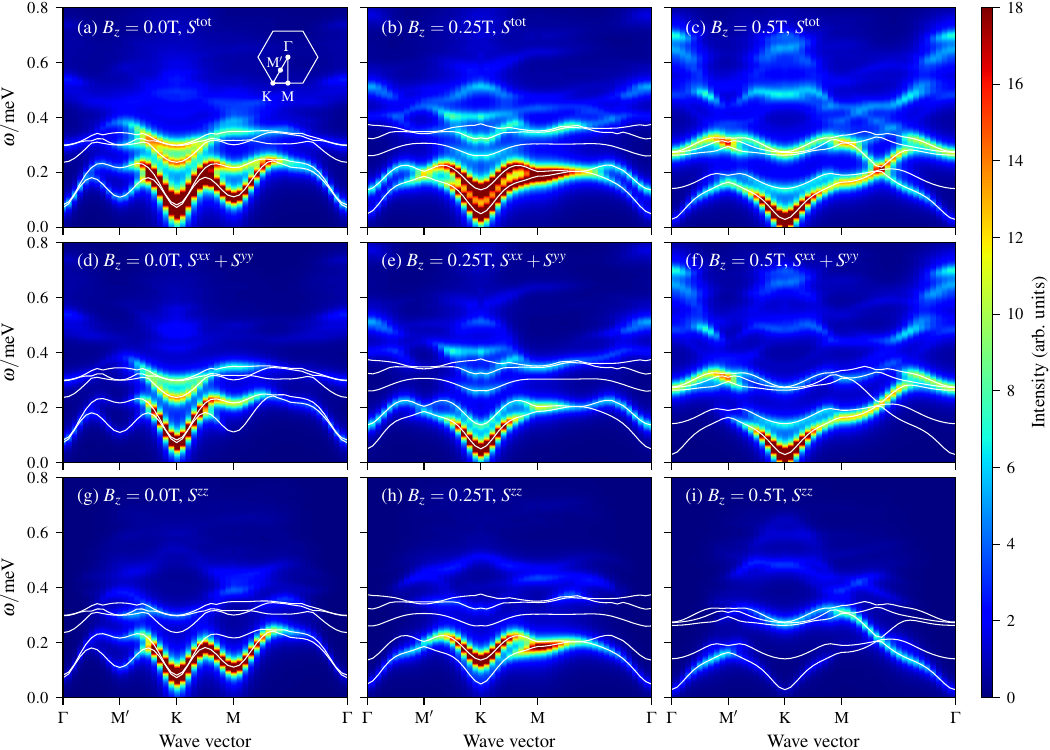}
\caption{Dynamical structure factor in the spin-supersolid Y phase of the TL-XXZ model with the parameters of $\mathrm{K_2Co(SeO_3)_2}$, computed using ET-CTMRG with $D = 4$ and $\chi = 100$ and shown along the high-symmetry momentum path $\Gamma$KM$\Gamma$ [inset panel (a)]. (a–c) Total spectral function, $S^{\mathrm{tot}} (\mathbf{k},\omega)$, at applied fields equivalent to $B_z = 0$, 0.25, and 0.5~T, respectively. (d–f) Transverse components, $S^{xx}(\mathbf{k},\omega) + S^{yy}(\mathbf{k},\omega)$, at the same three fields. (g–i) Longitudinal components, $S^{zz}(\mathbf{k},\omega)$, at the same three fields. White lines are guides to the eye showing the locations of the lowest five branches in the iPEPS spectrum.} 
\label{fig:KCoSeO_iPEPS_spectra}
\end{figure*}

\section{Spectral function of the TL-XXZ model near the Ising limit}
\label{secs4}

The fundamental quantity calculated in theory and numerics is the total spectral function 
\begin{equation}
S^{\mathrm{tot}} (\mathbf{k},\omega) = S^{xx} (\mathbf{k},\omega) + S^{yy} (\mathbf{k},\omega) + S^{zz}(\mathbf{k},\omega).
\label{eq:dsf}
\end{equation}
In Eq.~(9) of the main text we noted that this differs from the quantity measured in an INS experiment, which is weighted by the components of the $g$-tensor. The capability of iPEPS spectral calculations to separate $S^{\mathrm{tot}}$ into its transverse [$S^{xx}(\mathbf{k},\omega) + S^{yy}(\mathbf{k},\omega)$] and longitudinal [$S^{zz}(\mathbf{k},\omega)$] components is of particular advantage both in anisotropic systems and in an applied magnetic field. Thus the fact that the dynamical structure factor of K$_2$Co(SeO$_3$)$_2$ is dominated by $S^{zz}(\mathbf{k},\omega)$ means that additional information can be obtained from the channel decomposition. We recall that in both the INS experiment and our calculations, the field was applied in the direction of the $z$ component of spin space, which makes the definition of the terms transverse and longitudinal unambiguous; a different response would be found at finite fields applied in the plane of the TL.

In Figs.~\ref{fig:KCoSeO_iPEPS_spectra}(a-c) we show $S^{\mathrm{tot}} (\mathbf{k},\omega)$ for the three magnetic fields in the supersolid Y phase applied in recent INS experiments \cite{kcoseo_3}. Figures \ref{fig:KCoSeO_iPEPS_spectra}(d-f) show the transverse components, which are strongly suppressed in the INS data, and Figs.~\ref{fig:KCoSeO_iPEPS_spectra}(g–i) show the longitudinal channel, which reproduces all the primary features of the measured intensity. Quite generally, the zero-field spectra in Fig.~\ref{fig:KCoSeO_iPEPS_spectra} and Fig.~5 of the main text show minimal spectral weight around the $\Gamma$ point and maximal weight around K, while small finite fields cause some weight to be transferred to $\Gamma$. At zero field, the lowest-lying branches in each sector are nearly degenerate around K, their separation becoming visible beyond 1/3 of the way across the Brillouin zone, where the longitudinal mode is the lowest branch and the transverse one lies higher. The transverse branch shows a significantly weaker dip in energy (``roton'') than the longitudinal one at M, and by symmetry at the M$^\prime$ point mid-way between $\Gamma$ and K. Very recent polarized INS measurements \cite{zhu2026}, which achieve a clear separation of the transverse and longitudinal channels, indeed observe the weak transverse contribution lying higher at the M point. The second set of discrete levels, dispersing around 0.3~meV for the energy units of K$_2$Co(SeO$_3$)$_2$, is predominantly transverse and barely visible in experiment. Above these in the spectrum, all higher-lying features are significantly more diffuse in nature and very weak in intensity. 

Moving to finite fields, at 0.25~T one observes that the longitudinal branch is shifted up around K, in contrast to the transverse branch, whose energy remains nearly unchanged, marking this latter branch as the Goldstone mode of the supersolid phase. Around the $\Gamma$ point, it is the transverse mode that is pushed higher and gains some spectral weight. The two-magnon spectral weight clearly splits, and shifts predominantly upwards, remaining fully transverse in nature. By 0.5~T one finds a nontrivial reorganization of the spectrum, with the Goldstone mode staying the same at K, but its branch mixing into the two-magnon sector and showing continuous intensity across to the $\Gamma$ point. The longitudinal intensity has almost vanished at K, and remains only in some weaker branches that mix between the formerly clear one- and two-magnon sectors. The higher-lying transverse modes now provide both concentrated and diffuse features in the formerly continuum-like sector of the spectrum, and these are strong enough to become observable in experiment (Fig.~5 of the main text). 

\begin{figure}[t]
\centering
\includegraphics[width=0.72\columnwidth]{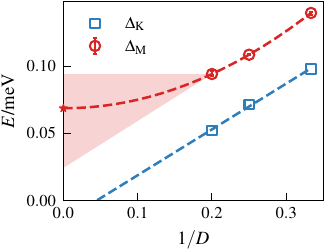}
\caption{Energy gaps at the K and M points in the $S = 1/2$ TL-XXZ model at applied field $B = 0$, shown as functions of $1/D$. Error bars for $\Delta_{{\rm M}}$ represent the variation between M points, as the iPEPS Ansatz we use does not preserve the $C_3$ symmetry that makes these equivalent. The shaded region represents the absolute bounds on the possible extrapolated values of the finite mode energy $\Delta_{\rm M}$.}
\label{fig:scaling}
\end{figure}

We remark again on the fact that our spectra have a finite-$D$ gap and hence do not capture the Goldstone mode dipping to zero energy at the K point. Previous studies of ordered magnetic states by the single-mode iPEPS Ansatz~\cite{Vanderstraeten2015, Vanderstraeten2019, ponsioen2020, tri_afm_1, tri_afm_2} have already documented clearly that the gap in the iPEPS spectrum converges to 0 as $1/D$, redeeming the expected physics, and this has also been shown for a gapless quantum spin liquid \cite{kagome_sl}. In Fig.~\ref{fig:scaling} we show the mode energies at K and M ($\equiv$ M$^\prime$) for the lowest branches at zero field in the K$_2$Co(SeO$_3$)$_2$ model, using the accessible system sizes $D = 3$, 4, and 5. Whereas the Goldstone-mode energy at K shows $1/D$ convergence to 0, the energy retains a robust finite value at the dips characterizing the M points. Here we note that, unlike for the Goldstone-mode energy, a systematic functional form in $1/D$ for a quantity extrapolating to a finite energy is not known. While it would be desirable to map this functional form to larger $D$ values, $D = 6$ is already at our current computational limits and would require some weeks of GPU time. 

Nevertheless, the very crude error-bounding procedure sketched in Fig.~\ref{fig:scaling} suggests a mode energy at the M-point dip in the vicinity of 0.06~meV. This value is approximately half of the energy at which the maximum in the INS intensity appears in Fig.~5(d) of the main text (from Fig.~5 of Ref.~\cite{kcoseo_3}). Taken together with the fact that these authors' own QMC calculations returned an energy of approximately 0.08~meV, this result suggests that the parameters of the model (specifically, the ratio $J_{xy}/J_z = 0.07$), which were determined from thermodynamic measurements \cite{kcoseo_2,kcoseo_3,kcoseo_4}, could be further refined by a more detailed modelling of the spectroscopic data. 

From a theoretical standpoint, some authors have discussed the possibility of fractional excitations in the triangular-lattice Ising antiferromagnet \cite{Jia2024}, and early INS measurements reported continuum-like spectral features \cite{kcoseo_2}. It is clear that local single-spin flips in an ordered configuration incur an energy penalty in an Ising model, but nonlocal processes constructed in the highly degenerate ground manifold of the triangular-lattice model may provide the possibility of fractionalization behavior near the Ising limit. The proposal that the ground state is proximate to a U(1) DSL \cite{Jia2024} mandates that the gapless spin response at the $\Gamma$, K, and M points arising from the spinons of the U(1) DSL should become gapped in the presence of the combined Ising and in-plane order of the spin supersolid. These authors then invoke a separate magnon sector to account for the Goldstone mode of the supersolid. 

In reality, the entire spin response must be described using a single set of spin degrees of freedom, either spinons or magnons, and the INS and iPEPS spectra show how the strong quasiparticle interactions in either framework are manifest. Certainly the lowest, magnon-like branches have pronounced dips at the M points, giving dispersions very far from those of linear spin-wave theory, even if they are clearly gapped at these points. The U(1) DSL interpretation may indeed offer a framework in which the discrete features ascribed to a one- and a two-magnon response are strongly bound, low-energy states of spinons that can appear as nearly-free quasiparticles at higher energies. Nevertheless, our iPEPS spectra show little evidence for spinon-like scattering at any low energies, and are more readily interpreted in this regime as a system of $\Delta S = 1$  excitations with significant INS resolution effects and background scattering \cite{tri_afm_2}. In this way our iPEPS spectra provide the benchmark for an analytical description of the interacting quantum many-body system and raise the bar for justifying claims of truly exotic states. 

We note in closing that the supersolid Y phase occupies only a very narrow region of the field-induced phase diagram for a system as close to the Ising limit as K$_2$Co(SeO$_3$)$_2$. For small $J_{xy}/J_z$, the phase diagram below saturation is dominated by the 1/3-plateau or UUD (up-up-down) phase, which is a simple product state with minimal additional correlation effects. The V phase appearing between the UUD and fully polarized phases has similar supersolid properties to the Y phase; we did not perform iPEPS calculations for the V phase because there are no INS data with which to compare. For the parameters of K$_2$Co(SeO$_3$)$_2$ ($J_{xy}/J_z = 0.07$), the Y phase appears from 0 to approximately 0.8~T \cite{kcoseo_2} and the V phase from approximately 20 to 21~T \cite{kcoseo_4}.

\end{document}